\documentclass[aps,prx,twocolumn,longbibliography,superscriptaddress,floatfix,nofootinbib]{revtex4-2}

\usepackage{times}

\usepackage[english]{babel}
\usepackage{latexsym}
\usepackage{graphics}
\usepackage{subfigure}
\usepackage{epsfig}
\usepackage{color}
\usepackage{hyperref}
\usepackage{braket} 
\usepackage[T1]{fontenc}
\usepackage[latin9]{inputenc}
\usepackage{amstext}
\usepackage{amssymb}
\usepackage{graphicx}
\usepackage{braket}
\usepackage{babel}
\usepackage{amsmath}
\usepackage{gensymb}
\usepackage{cancel}

\hypersetup{
colorlinks=true,
citecolor=blue,
linkcolor=red,
urlcolor=black
}


\newcommand{\ie}{i.e.}

\newcommand{\Er}{E_{\textrm{r}}}

\newcommand{\kB}{k_{\textrm{\tiny {B}}}}
\newcommand{\lambdadB}{\lambda_{\tiny T}}

\newcommand{\aOneD}{a_{\textrm{\tiny 1D}}}

\begin{document}

\title{
Observation of anyonic thermodynamics and generalized Pauli principle
}

\affiliation{State Key Laboratory of Photonics and Communications, School of Electronics, Peking University, Beijing 100871, China}
\affiliation{Institute of Carbon-based Thin Film Electronics, Peking University, Shanxi, Taiyuan 030012, China}
\affiliation{Center for Nonlinear Phenomena and Complex Systems, Universit\'e Libre de Bruxelles, CP 231, Campus Plaine, 1050 Brussels, Belgium}

\author{Fansu Wei}
\homepage{These authors contributed equally to this work.}
\affiliation{State Key Laboratory of Photonics and Communications, School of Electronics, Peking University, Beijing 100871, China}
\author{Chi Zhang}
\homepage{These authors contributed equally to this work.}
\affiliation{State Key Laboratory of Photonics and Communications, School of Electronics, Peking University, Beijing 100871, China}
\author{Zimeng Ye}
\affiliation{State Key Laboratory of Photonics and Communications, School of Electronics, Peking University, Beijing 100871, China}
\author{Dengbo Wang}
\affiliation{Institute of Carbon-based Thin Film Electronics, Peking University, Shanxi, Taiyuan 030012, China}
\author{Botao Wang}
\email{botao.wang@ulb.be}
\affiliation{Center for Nonlinear Phenomena and Complex Systems, Universit\'e Libre de Bruxelles, CP 231, Campus Plaine, 1050 Brussels, Belgium}
\author{Xiaoji Zhou}
\email{xjzhou@pku.edu.cn}
\affiliation{State Key Laboratory of Photonics and Communications, School of Electronics, Peking University, Beijing 100871, China}
\affiliation{Institute of Carbon-based Thin Film Electronics, Peking University, Shanxi, Taiyuan 030012, China}
\author{Hepeng Yao}
\email{hepeng.yao@pku.edu.cn}
\affiliation{State Key Laboratory of Photonics and Communications, School of Electronics, Peking University, Beijing 100871, China}

\date{\today}

\begin{abstract}


Anyons are quasiparticles with quantum statistics interpolating between those of bosons and fermions. Two distinct manifestations of anyonic behaviour have been theoretically established: fractional exchange statistics where particle exchange can produce any phase, and generalized exclusion statistics which extends the Pauli exclusion principle. While anyons exhibiting fractional exchange statistics have been observed in diverse platforms, experimental realizations of generalized exclusion statistics and direct measurements of its thermodynamic signatures have remained elusive.
Here, we realize an anyonic thermodynamic ensemble obeying generalized exclusion statistics and detect its anyonic thermodynamics in a one-dimensional strongly interacting quantum gas. To achieve this, we exploit the bijective mapping between dynamical and statistical interactions in one dimension. By tuning interaction strength and temperature over a wide range, we measure the equation of state and identify clear departures from Bose-Einstein and Fermi-Dirac statistics. These deviations are quantitatively captured by generalized exclusion statistics, providing direct evidence for the generalized Pauli principle. Independent probes of other thermodynamic quantities including pressure and the Tan contact further validate this framework. Our results establish a versatile platform for engineering anyonic exclusion statistics and open the door to thermodynamic applications of anyons in quantum technologies.


\end{abstract}

\maketitle

Elementary particles in nature are fundamentally classified as either bosons or fermions. Under exchange, bosonic wave functions acquire a phase factor $+1$ (symmetric), while fermionic ones acquire $-1$ (antisymmetric). Such a symmetry requirement has profound physical consequences. Fermions obey the Pauli exclusion principle and repel each other, 
forming the building blocks of matter and the periodic table. By contrast, bosons can macroscopically occupy a single quantum state,  
mediating fundamental interactions and giving rise to intriguing phenomena such as Bose-Einstein condensation.
In low-dimensional quantum systems, however, the possibility of picking up any phase $e^{i\theta}$ that corresponds to quasiparticles called anyons~\cite{leinaas1977, Greiter2024, PhysRevLett.49.957}, has greatly enriched the landscape of quantum statistics.
In two dimensions (2D), anyons emerge as elementary excitations of topologically ordered states~\cite{PhysRevLett.52.1583, PhysRevLett.53.722, feldman2021}. They obey fractional exchange statistics (FES) interpolating between bosons and fermions, and hold the promise for the realization of topological quantum computing~\cite{KITAEV20032, RevModPhys.80.1083}. Recent experiments have provided direct evidence of FES across diverse platforms, including solid-state materials~\cite{feldman2021}, superconducting circuits~\cite{PhysRevLett.117.110501,Xu2024} and trapped ions~\cite{Iqbal2024}.

\begin{figure*}
    \centering
    \includegraphics[width=1\linewidth]{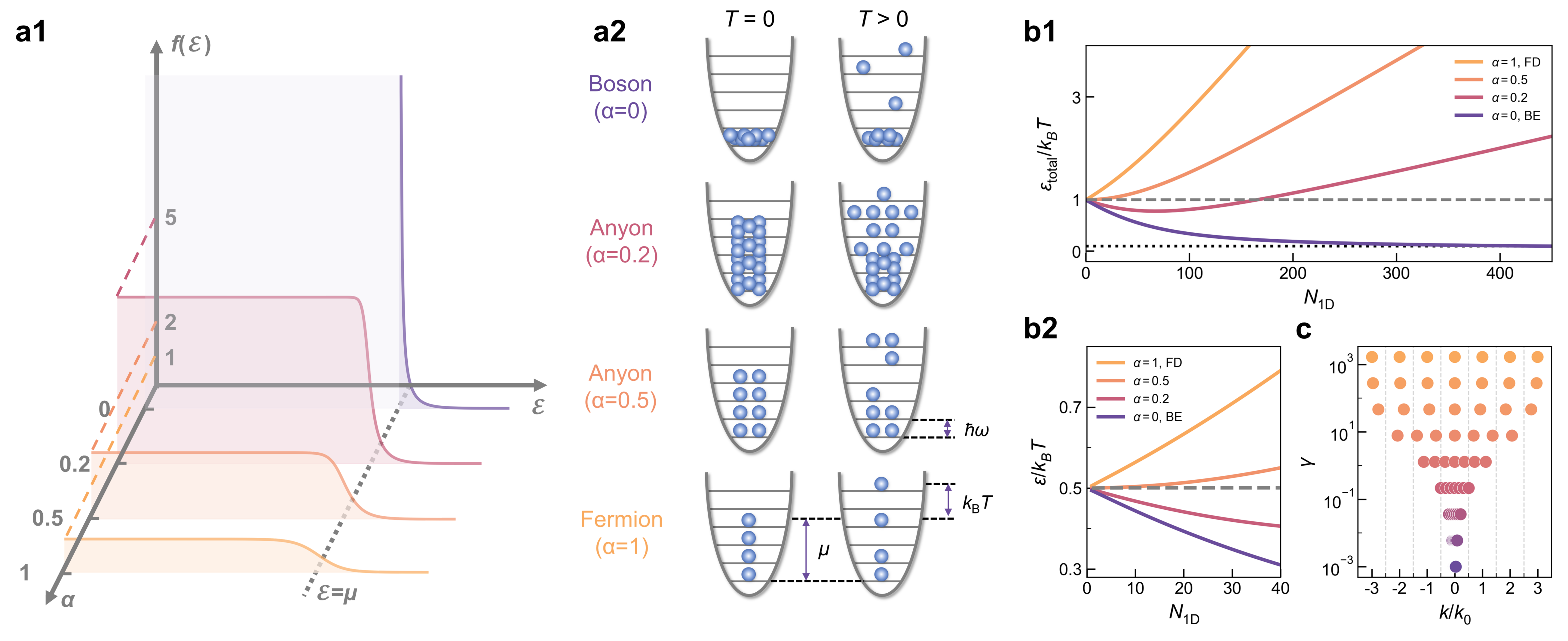}
    \caption{\textbf{Illustration of the anyonic generalized exclusion statistics.}
    \textbf{a1,} Mean occupation number $f(\mathcal{E})$ versus state energy $\mathcal{E}$ at finite temperature for generalized exclusion statistics with different statistical parameter $\alpha$. The curves represent Bose-Einstein (BE, $\alpha=0$, purple), Fermi-Dirac (FD, $\alpha=1$, orange), and two typical anyonic ($\alpha=0.2$, light purple; $\alpha=0.5$, dark orange) distributions. $\mu$ denotes the chemical potential. At high energies ($\mathcal{E} \gg \mu$), all distributions asymptotically approach zero. At lower energies, $f(\mathcal{E})$ increases and saturates at the statistical upper bound $1/\alpha$, consistent with the generalized Pauli exclusion principle.
    \textbf{a2,} Schematic of state occupations in a 1D harmonic trap with trap frequency $\omega$, at both zero and finite temperatures for various $\alpha$. Energy levels are equally spaced by $\hbar\omega$.
    \textbf{b1,} Total energy per particle scaled by temperature $\varepsilon_{\rm total}/(\kB T)$, as a function of the particle number $N_{\rm 1D}$ in a 1D harmonic trap. In the low-density limit, all gases converge to the classical Maxwell-Boltzmann behavior $\varepsilon_{\rm total} = \kB T$ (gray dashed line). As $N_{\rm 1D}$ increases, the BE gas (purple solid line) asymptotically approaches the ground-state energy $\hbar\omega/2$ (black dotted line), whereas the FD gas (orange solid line) exhibits a monotonic increase due to Pauli exclusion. The anyonic cases ($\alpha=0.2$ and $\alpha=0.5$) display intermediate behaviors.
    \textbf{b2,} Close-up view of the scaled internal energy per particle $\varepsilon/(\kB T)$, obtained by subtracting the external trapping potential energy from $\varepsilon_{\rm total}$. The plotted range of $N_{\rm 1D}$, from 0 to 40, corresponds to the experimental parameter regime.
    \textbf{c,} 
    Quasi-momentum distribution of the Lieb-Liniger gas for varying interaction strength $\gamma$, normalized by $k_0 = 2\pi/L$ with $L$ the system size.
    As $\gamma$ increases, the quasi-momentum distribution evolves from bosonic clustering ($\gamma \to 0$) to uniform fermionic spacing ($\gamma \gg 1$), directly visualizing the emergence of generalized Pauli exclusion.
    }
    \label{fig:1}
\end{figure*}

There are two known frameworks of quantum statistics for anyons~\cite{Greiter2024}.
Complementary to the fractional \textit{exchange} statistics, generalized \textit{exclusion} statistics (GES), introduced by Haldane as an extension of the Pauli exclusion principle~\cite{Haldane1991}, have provided a compelling formulation for studying anyons from the thermodynamic perspective. 
GES describes systems in which the dimension of the state space changes linearly with particle numbers, allowing partial occupation of a single state~\cite{Haldane1991}.
As a consequence, the equilibrium occupation number $f(\mathcal{E})$ for a state of energy $\mathcal{E}$ takes the form~\cite{GES}
\begin{align}
    f(\mathcal{E}) &= \frac{1}{w(\zeta) + \alpha},\label{Eq:GES1} \\
    w(\zeta)^{\alpha} (1+w(\zeta))^{1-\alpha} &= \zeta = \exp \left( \frac{\mathcal{E}-\mu}{k_{\mathrm{B}} T} \right),\label{Eq:GES2}
\end{align}
with $\mu$ the chemical potential, $T$ the temperature and $\kB$ the Boltzmann constant. 
The parameter $\alpha$ characterizes the statistics: $\alpha=0$ and $\alpha=1$ recover the Bose-Einstein (BE) and Fermi-Dirac (FD) distributions, respectively. For intermediate values, particles exhibit anyonic characteristics with a maximum occupation number of $1/\alpha$ (Fig.~\ref{fig:1}a1). 
In previous studies, anyonic excitations obeying GES have been identified in several 1D models theoretically, including the Haldane-Shastry~\cite{Haldane1988,Shastry1988} and Calogero-Sutherland models~\cite{Ha1994}, accompanied by interesting thermodynamical properties~\cite{GES, PhysRevLett.73.3331, PhysRevLett.74.1493}. 
However, because these Hamiltonians contain an inverse-square term, their experimental realization remains challenging.  

In recent years, low-dimensional cold atom systems have emerged as a promising platform for engineering anyonic behaviors~\cite{Frolian2022,anyongreiner,Dhar2025}. While recent experiments have revealed signatures of FES through quantum walks~\cite{anyongreiner} and impurity correlations~\cite{Dhar2025}, realization of an anyonic thermodynamic ensemble with GES statistics remains a major challenge. The unique feature of 1D may provide a direct route: because collision is the only way to interchange two particles in 1D, dynamical interaction can be mapped bijectively onto statistical exclusive interaction~\cite{GES,Guan1,Guan3}. A paradigmatic example is strongly interacting bosons, known as Tonks-Girardeau gas~\cite{girardeau1960, Paredes2004, kinoshita2004}, whose thermodynamic properties map onto those of ideal fermions. Upon reducing interaction strength or increasing temperature, the fermionization gradually breaks down, resulting in a smooth crossover back to bosonic behavior~\cite{yao-tancontact-2018}. Such continuous transitions may therefore provide a natural setting for exploring intermediate statistics with anyonic thermodynamic properties.

Here we report the observation of anyonic thermodynamics governed by GES using 1D strongly interacting $^{6}\rm Li_2$ molecules.
By tuning the interplay between temperatures and interactions, we measure the equation of state and realize a continuous crossover from fermionic to bosonic statistics. Our measurements of the energy-particle number relation reveal equilibrium states quantitatively characterized by the GES parameter $\alpha$.
As $\alpha$ continuously evolves from $1$ to $0$, we show evidence of the generalized Pauli exclusion principle with a maximum occupation number $1/\alpha$ exceeding unity. The underlying mechanism is captured within the framework of the thermal Bethe ansatz.
Extending to other thermodynamic quantities, we further validate this description using independent probes of pressure and the Tan contact.
Our results establish a versatile experimental platform for engineering fractional statistics and open new avenues for investigating the thermodynamics of anyonic matter.

\begin{figure*}
    \centering
    \includegraphics[width=1\linewidth]{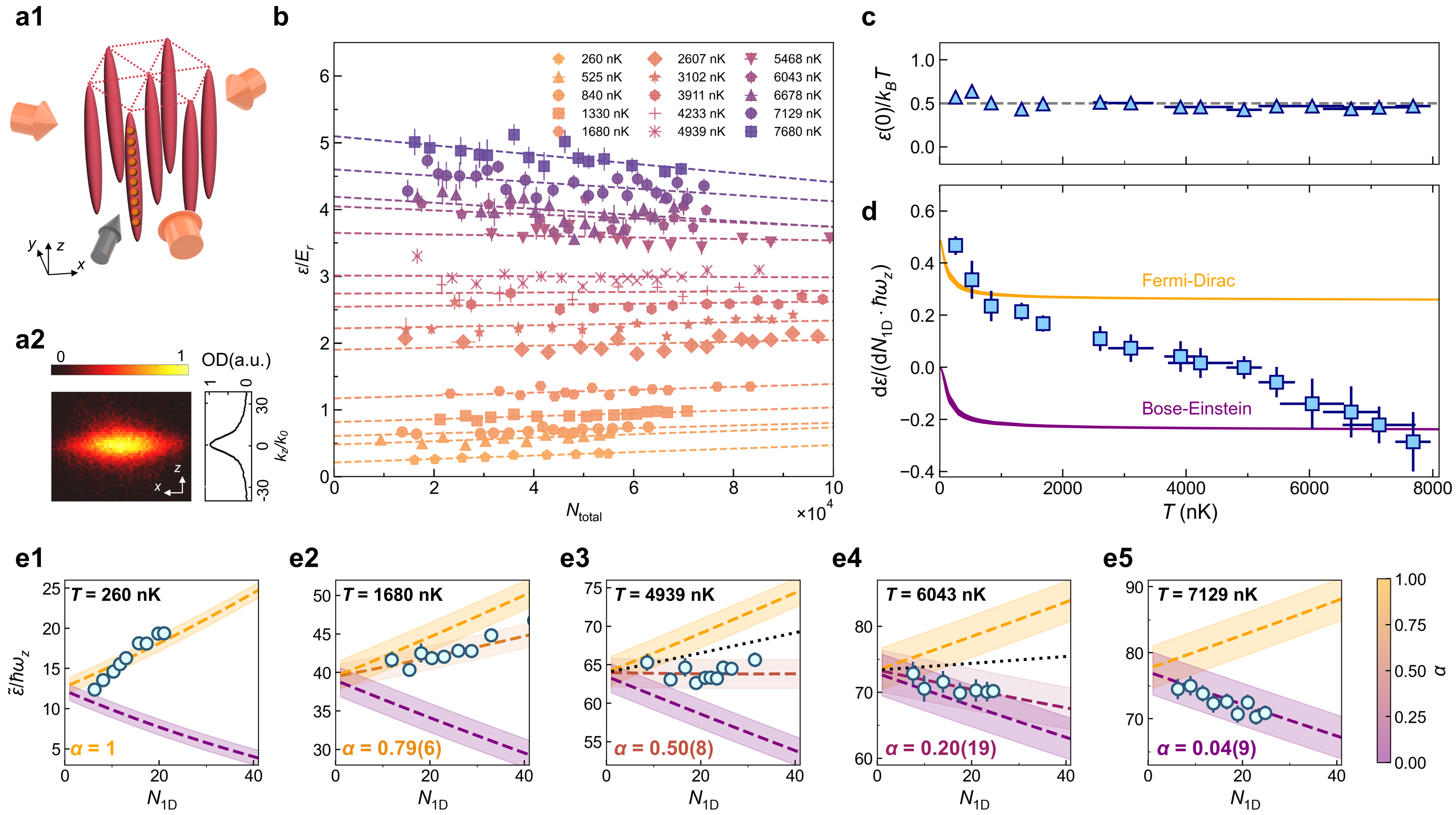}
    \caption{
    \textbf{Observation of the equation of state for anyonic thermodynamics.}
    \textbf{a1,} Experimental realization of 1D Bose molecular gases in tubes formed by 2D triangular optical lattices. The lattice is created by three coherent traveling-wave laser beams arranged at $120\degree$ to each other (yellow arrow).
    Imaging is performed perpendicular to the tubes, at $30\degree$ to one lattice beam path (grey arrow). 
    \textbf{a2,} Typical time-of-flight absorption images for $T=260(10)\ \mathrm{nK}$ and $a_s = 743a_0$. The right panel shows the obtained momentum distribution $n(k_z)$ along the $z$-direction. 
    \textbf{b,} The energy per particle $\varepsilon$ (in units of the recoil energy $ \Er $) as a function of total particle number $N_{\rm total}$ for different temperatures $T$ from $260(10)$ nK to $7680(287)$ nK, each represented by a distinct color. Data points are means of at least 12 independent measurements and the error bars represent the standard error of $N_\textrm{total}$ and $\varepsilon$.
    Dashed lines are linear fits per temperature dataset. 
    \textbf{c,} 
    The intercept $\varepsilon(0)/\kB T$ of the linear fit in (\textbf{b}) as a function of temperature $T$.
    The black solid line indicates the Maxwell-Boltzmann limit $\varepsilon=\kB T/2$.
    \textbf{d,} Average energy derivative $\mathrm{d} \varepsilon / {\mathrm{d} N_{\rm 1D}}$ versus $T$. Blue squares represent slopes extracted from the linear fits in (\textbf{b}), with $N_{\rm total}$ mapped to $N_{\rm 1D}$ in a single tube. 
    Error bars incorporate uncertainties from both linear fits and temperature measurements. 
    Orange and purple lines are theoretical predictions of FD and BE distributions for an equivalent 1D system under the same conditions. 
    \textbf{e1-e5}, Example datasets at five representative temperatures $T=260(10)$, $1680(80)$, $4939(287)$, $6043(520)$, $7129(592)$ nK. 
    For better visualizing the comparison, we rescale $\tilde{\varepsilon}$ by matching the intercept to $\kB T/2$.
    Blue circles are experimental data points. Analytical GES curves are shown for fermions (orange dashed lines), bosons (purple dashed lines) and anyons (dashed lines color-coded by $\alpha$). Shaded bands indicate confidence intervals arising from experimental temperature uncertainty. Black dotted lines represent SU(N) Fermions for comparison. 
    }
    \label{fig:2}
    
\end{figure*}


The experimental system reported here is confined in a harmonic potential with trap frequency $\omega$. Its eigenenergy reads $E_j=(j+1/2)\hbar\omega$ with 
$\hbar$ the reduced Planck constant, see Fig.~\ref{fig:1}a2. According to this filling pattern, the total energy per particle $\varepsilon_{\rm total}$ and its related equation of state can serve as a good indicator for determining $\alpha$~\cite{Guan2}.
In Fig.~\ref{fig:1}b, we plot $\varepsilon_{\rm total}$ as a function of 1D particle number $N_{\rm 1D}$ for ideal gases under different statistical distributions based on Eqs.~(\ref{Eq:GES1}, \ref{Eq:GES2}).
In the dilute limit $N_{\rm 1D} \to 0$, the identical particle postulate of quantum mechanics fails and all types of statistics collapse into the classical Maxwell-Boltzmann limit. According to the equi-partition theorem, 
it reads $\varepsilon_{\rm total}=\varepsilon+\varepsilon_V=\kB T$ (black dashed line) with $\varepsilon$ and $\varepsilon_V$ the averaged internal and external potential energies, respectively. 

By increasing $N_{\rm 1D}$, the indistinguishability becomes substantial and the curves deviate. 
For FD statistics ($\alpha=1$, orange), each additional particle must occupy a higher state, 
leading to a linearly increasing dependence $\varepsilon_{\rm total} \propto N_{\rm 1D}$. 
In contrast, BE statistics ($\alpha=0$, dark purple) allow macroscopic occupation of the ground state $\lim_{N\rightarrow\infty} \varepsilon_{\rm total}=\hbar \omega /2$.
Since $\kB T \gg \hbar \omega /2$ usually holds in experiments, $\varepsilon_{\rm total}$ should decrease from $\kB T$ and saturates at $\hbar \omega /2$.
For the anyonic cases, $\alpha$ lies between these two extremes. Here, we show two examples at $\alpha=0.5$ (dark orange) and $\alpha=0.2$ (purple).
As the internal energy $\varepsilon=\varepsilon_{\rm total}-\varepsilon_V$ can be readily measured in cold-atom experiments~\cite{Minguzzi_1997, PhysRevLett.132.243403, PhysRevLett.95.250404, localization2025}, we plot the $\varepsilon-N_\textrm{1D}$ curve within the experimentally accessible regime in Fig.~\ref{fig:1}b2.
Clearly, these curves exhibit different monotonic behaviors starting from $\kB T/2$, suggesting a good window for observing GES.


The intuition for GES arises from the bijective mapping between dynamical and statistical interactions in the Lieb-Liniger model of 1D interacting bosons~\cite{GES, Guan2, PhysRevLett.74.1493}.
It can be solved exactly by Bethe ansatz at zero temperature, whose solution in quasi-momentum $k$ space is shown in Fig.~\ref{fig:1}c~\cite{Guan3}.
As the Lieb-Liniger interaction parameter $\gamma$ increases, the roots gradually spread from a bosonic cluster near $k=0$ into a fermionic uniform distribution. 
Along such state-space evolution, GES naturally appears (see detailed derivations in Supplementary Information).
As discussed in Ref.~\cite{yao-tancontact-2018}, increasing temperature will also equivalently suppress the Pauli exclusion of dynamical interactions and thus lead to anyonization. Below, we prepare the anyonic thermodynamic ensemble based on this picture.

Our experiment starts from preparing $^{6}\rm Li_2$ Feshbach molecules composed of atoms in the two lowest hyperfine states in a cross-beam dipole traps~\cite{jochim2003, PhysRevA.109.043313}. The inter-molecule interaction strength can be set over a range by tuning the $s$-wave scattering length $a_{s}$ between molecules via the magnetic Feshbach resonance. 
The total particle number $N_{\rm total}$ and temperature $T$ are finely controlled through precise adjustment of the cooling sequence,
similarly as Refs.~\cite{Anomalouscooling,Tian2026}. Then, a 2D triangular optical lattice with wavelength $\lambda=1064$ nm in the horizontal $xy$-plane is adiabatically ramped up to $V_{\rm 2D}=10 \Er $ within 40 ms, with $ \Er  = 2\pi^2\hbar^2/(\mathit{m}\lambda^2)$ the recoil energy and $m$ the mass of a $^{6}\rm Li_2$ molecule.
As illustrated in Fig.~\ref{fig:2}a1, we create approximately 3000 isolated 1D tubes along the $z$ direction with longitudinal trap frequency $\omega_z/2\pi \simeq213$ Hz. The 3D scattering length is set to $a_s = 743a_0$, giving the 1D scattering length $|\aOneD|=2831.6a_0$ and $\gamma\geq20$, with $a_0$ the Bohr radius.
After reaching thermal equilibrium, we carry out measurements
via the time-of-flight (TOF) process. The optical potentials are switched off and the particles are detected by absorption image after a typical 2.5 ms expansion (Fig.~\ref{fig:2}a2).
With the obtained momentum distribution $n(k_z)$ along $z$ direction, we can extract the total particle number $N_{\rm total}=\int n(k_z) \mathrm{d}k_z$ and total internal energy $E=\int n(k_z) (k_z^2/2m)\mathrm{d}k_z$~\cite{PhysRevLett.132.243403,PhysRevLett.95.250404, localization2025}. Also, we use the thermometry method in Refs.~\cite{Anomalouscooling,localization2025} to measure the temperature of 1D systems. The minimum temperature is $T=260$ nK, which is deeply in the Tonks-Girardeau regime~\cite{PhysRevLett.91.040403}. More experimental details can be found in the Methods and Supplementary Information.


From the TOF measurement, we extract the equation of state at different temperatures.
In Fig.~\ref{fig:2}b, we show the detected energy per particle $\varepsilon=E/N_{\rm total}$ as a function of the total particle number $N_{\rm total}$ for temperatures from $260(10)$ nK to $7680(287)$ nK (colored symbols from orange to dark purple).  
At the lowest temperature $T=260(10)$ nK, $\varepsilon$ increases with $N_{\rm total}$ at the greatest slope. As the temperature increases, while the absolute values of $\varepsilon$ increase, the slope decreases and eventually tilts toward the opposite direction, reaching its lowest minimum at $T=7680(287)$ nK. 
According to Fig.~\ref{fig:1}b2, this provides a strong hint that the statistics evolve continuously from FD towards BE.

To achieve a more quantitative understanding, we map the total particle number $N_{\rm total}$ to the equivalent 1D particle number $ N_{\rm 1D}$ (see Refs.~\cite{PhysRevLett.115.085301, Anomalouscooling, Guo2024} and Methods), and apply a linear fit to the $\varepsilon - N_{\rm 1D}$ curves.
The obtained intercept and slope are shown in Fig.~\ref{fig:2}c and d, respectively. 
In Fig.~\ref{fig:2}c, we find the intercept $\varepsilon(0)=\lim_{N_{\rm 1D}\rightarrow 0} \varepsilon$ is always around $\kB T/2$ within $10.7 \%$, recovering the classical gas limit discussed in Fig.~\ref{fig:1}b2. 
More importantly, the slopes versus $T$ are shown in Fig.~\ref{fig:2}d. They are plotted with the unit of $\hbar \omega_z$ and compared with analytical curves of FD and BE distributions.
Below 800 nK, the experimental data sit around the FD prediction ($\mathrm{d} \varepsilon/\mathrm{d} N_{\rm 1D} \sim 0.3 \hbar \omega_z$). As the temperature increases, the slope decreases and eventually changes its sign to negative, finally approaching the BE limit ($\mathrm{d} \varepsilon / {\mathrm{d} N_{\rm 1D}} \sim -0.2\hbar \omega_z$) above 7000 nK. 

In Fig.~\ref{fig:2}e1-e5, we further show the detailed  $\varepsilon - N_{\rm 1D}$ data for five representative temperatures (blue circles). On top, we plot the analytical expression for GES at $\alpha=1$ (orange dashed line, FD) and $\alpha=0$ (purple dashed line, BE).
The shaded area reflects the uncertainties in experimental parameters. 
Clearly, at $T=260(10)$ nK (e1) and $7129(592)$ nK (e5), the experimental data agree well with the FD and BE distribution, respectively. 
For the intermediate temperatures $T=1680(80)$, $4939(287)$, and $6043(520)$ nK (e2-e4), the results lie in between, consistent with the anyonic scenario illustrated in Fig.~\ref{fig:1}b2. By comparing the experimental data with theoretical predictions of GES at various $\alpha$, 
we identify anyonic statistics with $\alpha=0.79(6)$, $0.50(8)$, $0.20(9)$. Here, the errors come from both the linear fit of the $\varepsilon-N_{\rm 1D}$ curve and the calibration of experimental parameters. 
The corresponding GES curves are shown as dashed lines. We note that all the experimental data within 800 nK $<T<7000$ nK fall in between, indicating $0<\alpha<1$ and a generalized Pauli exclusion with maximum occupancy per state $f_{max}(\mathcal{E})=1/\alpha$. For instance, at $T=4939(287)$ nK (e3), we recognize $\alpha=0.5$ which signatures semionic exclusive statistics with a maximum filling of $2$ per state. We further confirm that their behavior is different from that of ideal fermions with additional degeneracy, such as ideal SU(N) fermions (black dotted line in e3 and e4), thereby revealing the unique features of indistinguishability in our anyonic ensemble.

\begin{figure}
    \centering
    \includegraphics[width=0.95\linewidth]{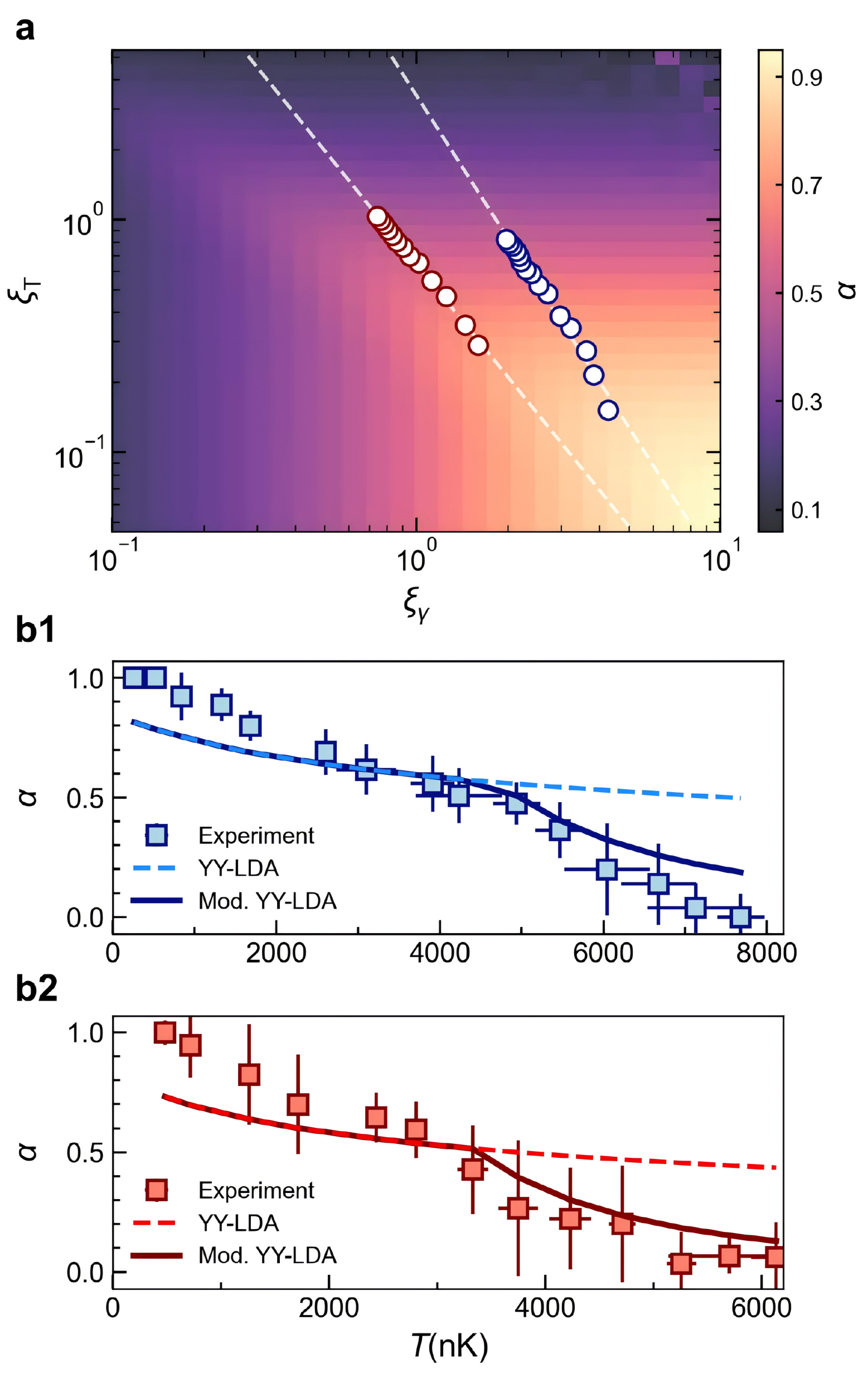}
    \caption{\textbf{Phase diagram of statistical parameter $\alpha$.}
    \textbf{a,} 
    Phase diagram of the statistical parameter $\alpha$ as a function of reduced temperature $\xi_T$ and reduced interaction $\xi_{\gamma}$, computed from YY-LDA.
    The color code indicates crossover from fermionic statistics ($\alpha=1$, orange) to bosonic statistics ($\alpha=0$, purple).  
    The parameter ranges for experimental measurements are indicated for $V_{\rm 2D}=10\Er$ (blue circles, measurement in Fig.~\ref{fig:2}b) and $V_{\rm 2D}=5\Er$ (red circles), with their linear fit on a log-log scale (dashed lines). 
    \textbf{b1,} The measured statistical parameters $\alpha$ as a function of temperature $T$ (squares), together with the corresponding theoretical predictions from (\textbf{a}) (dashed line). The solid line shows the modified $\tilde{\alpha}$ by considering the transverse excitations.
    \textbf{b2,} is the counterpart of \textbf{b1} but with $V_{\rm 2D}=5\Er$. 
    }
    \label{fig:3}
\end{figure}



To further elucidate the underlying mechanisms, we consider the Lieb-Liniger Hamiltonian
\begin{equation}
\hat{H} = \sum_i\left[-\frac{\hbar^2}{2m}\frac{\partial^2}{\partial x_i^2}+\frac{1}{2}m\omega^2 x_i^2\right]+g_{\rm 1D}\sum_{i<j}\delta(x_i-x_j),
\label{eq:H}
\end{equation}
with $g_{\rm 1D}=-2\hbar^2/m \aOneD$ the interaction coupling constant.
It is linked with the interaction parameter $\gamma=mg_{\rm 1D}/\hbar^2 n$ with $n$ the particle density~\cite{lieb1963a}.
At finite temperature, it can be solved by the Yang-Yang thermodynamics, also known as thermal Bethe ansatz~\cite{yang1969}. Combining this approach with local density approximation (YY-LDA), the thermodynamic properties of the trapped gas can be determined~\cite{yao-tancontact-2018, PhysRevLett.119.165701}. 
In principle, these properties depend on four independent parameters: the particle number $N_{\rm 1D}$, temperature $T$, trap frequency $\omega$ and interaction strength $g_{\rm 1D}$. 
As demonstrated in Ref.~\cite{yao-tancontact-2018}, 
the free variables of this problem reduce to two dimensionless parameters: the reduced temperature $\xi_T=-\aOneD/\lambda_T$ and the reduced interaction $\xi_{\gamma}=-a_{\rm ho}/\aOneD\sqrt{N_{\rm 1D}}$, with $a_{\rm ho}=\sqrt{\hbar/m\omega}$ the oscillation length and  $\lambda_T=\sqrt{2\pi\hbar^2/m\kB T}$ the de Broglie wavelength.



By computing the $\varepsilon-N_{\rm 1D}$ properties from YY-LDA, we can determine the statistical parameter $\alpha$ for trapped Lieb-Liniger gas over a wide parameter range (Methods), see Fig.~\ref{fig:3}a. This theoretical phase diagram confirms the anyonization picture above.
In the limit of strong interaction and low temperature ($\xi_T\ll1, \xi_{\gamma} \gg1$), the system is deeply in the fermionized regime with $\alpha\simeq1$ (light orange area).
By increasing temperatures or reducing interactions, the Pauli exclusion is gradually suppressed and
$\alpha$ decreases, eventually recovering the BE limit $\alpha=0$ (dark purple area). 
The parameter range of measurement in Fig.~\ref{fig:2} is marked by blue circles on this phase diagram.
They approximately fall along a straight line on the log-log scale, originating from the deep FD region and gradually moving into the BE area.

In Fig.~\ref{fig:3}b1, we further quantitatively compare the measured $\alpha$ values (blue squares) with the YY-LDA predictions (dashed line).
While both show a clear decreasing behavior with $T$, the experimental data decrease faster at high temperatures. 
This can be attributed to
additional degeneracy of the state induced by the transverse dimensions. 
Considering such additional state occupancy approximately (Methods), we plot the modified statistical parameter $\tilde{\alpha}$ and find it cures the discrepancy in the high-temperature limit (solid curve).
We also note that in the low-temperature limit, the experimental data reach $\alpha=1$ faster than YY-LDA, which may be caused by the tube distributions.
To further check the generality, we carry out another set of measurements with transverse lattice depth $V_{\rm 2D}=5\Er$, tracing a different parameter path in Fig.~\ref{fig:3}a (red circles). The measured $\alpha$ values, as well as their theoretical prediction, are shown in Fig.~\ref{fig:3}b2. Similarly, the experimental data are compatible with the YY-LDA results, and a clear FD-BE crossover is observed.

\begin{figure}
    \centering
    \includegraphics[width=0.9\linewidth]{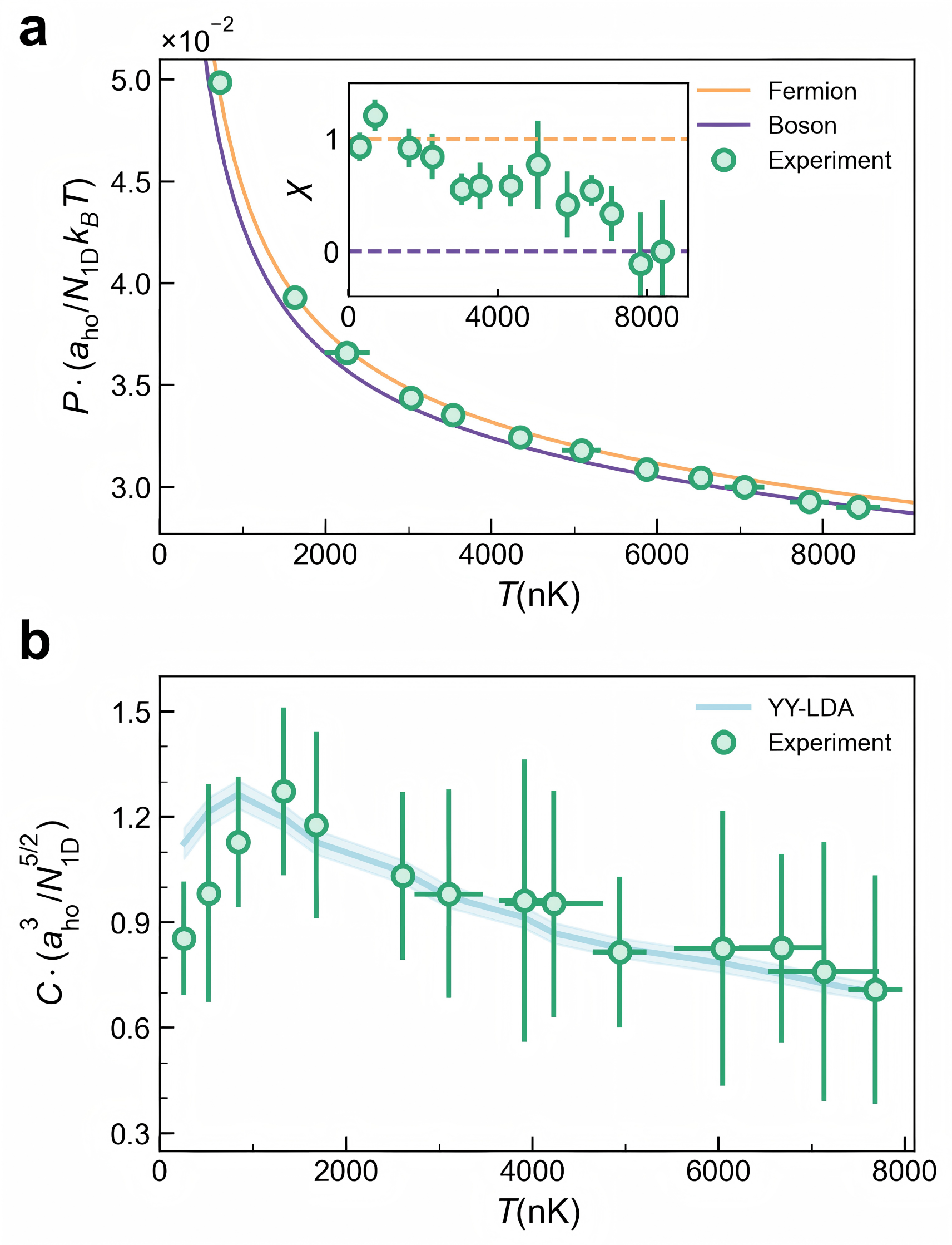}
    \caption{\textbf{Measurement of anyonization in other thermodynamic properties.}
    \textbf{a,} Pressure $P$ as a function of temperature $T$ (green circles), together with the analytical curves of FD  and BE statistics (orange and purple solid lines). The inset shows the metric $\chi=(P_{\rm Exp}-P_{\rm B})/(P_{\rm F}-P_{\rm B})$ for evaluating the statistics. 
    \textbf{b,} The detected Tan's contact $C$ versus temperature.
    The green circles are experimental data obtained by taking the difference between the measurements at magnetic fields of 680 G ($\gamma\simeq20.4$) and 690 G ($\gamma\simeq23.4$), and the blue line is the YY-LDA result. All experimental results are averaged over 60 realizations.
    }
    \label{fig:4}
\end{figure}

To further verify that an anyonic thermodynamic ensemble is generated, we measure other thermodynamic quantities with independent probes.
We start with the measurement of thermodynamic pressure under fixed particle number $N_{\rm 1D}\approx12$ and other conditions the same as Fig.~\ref{fig:2}. By integrating \textit{in situ} absorption images, we measure the longitudinal density distribution $n(z)$. Then, the pressure $P$ can be obtained by~\cite{science.1214987, PhysRevLett.119.165701}
\begin{align}
    P=\int_{-\infty }^{\mu_0 } n(\mu )\mathrm{d} \mu=\int_{0}^{\infty }n(z)m\omega_z ^2\mathrm{d}z,
\end{align}
with $\mu_0$ the central chemical potential. 
The experimental result $P_\textrm{Exp}$ at different temperatures is shown in Fig.~\ref{fig:4}a.
Correspondingly, we also plot the analytical results of pressure for ideal bosons $P_{\rm B}$ (purple solid line) and fermions $P_{\rm F}$(orange solid line).
Although their difference is modest within the considered temperature range, the error bars of experimental data are much smaller than their difference and the statistical nature can still be identified.
To better view it, we introduce the parameter $\chi=(P_{\rm Exp}-P_{\rm B})/(P_{\rm F}-P_{\rm B})$ which quantifies the deviation from the two statistics (inset).
$\chi$ approaches 1(0) when the data are consistent with fermionic (bosonic) statistics.
Within the error bars, $\chi$ clearly evolves from unity to zero as temperature increases, indicating further evidence for anyonic thermodynamics.
Through the Maxwell relations based on energy and pressure, this finding implies that other thermodynamic quantities should likewise exhibit anyonic behaviors.

Another important thermodynamic quantity is the Tan contact $C$. It is the weight of the large-momentum tail and serves as the thermodynamic conjugate to $g_{\rm 1D}$~\cite{TAN20082952}. 
In the Lieb-Liniger model, the contact as a function of temperature $T$ reflects the effective statistics. 
It increases with $T$ in the fermionic regime and decreases with $T$ in the high-temperature bosonic regime~\cite{yao-tancontact-2018}.
Here, we measure the energy at two different interactions $a_s=743a_0$ and $854a_0$, and obtain the contact via the Tan sweep relation $\left.C=(4m/\hbar^2)\partial E_{\rm single}/\partial \aOneD\right|_{S}$~\cite{PhysRevLett.108.145305, Zou2021}, with $E_{\rm single}=N_{\rm 1D} \cdot \varepsilon$ the energy of an equivalent 1D tube.
The contact is measured at different temperatures for $N_{\rm 1D} \approx 20$ and the results are shown in Fig.~\ref{fig:4}b. 
Clearly, $C$ increases with $T$ until $T\simeq1500$ nK and then decreases, showing clear evidence of a Fermi-Bose crossover. This also agrees with the YY-LDA predictions (blue solid line). The measured maximum falls into the range of the anyonic regime 800 nK $<T<$ 7000 nK indicated in Fig.~\ref{fig:3}b. The corresponding crossover temperature reads $\xi_T=\aOneD/\lambda_T=0.4$, compatible with the prediction in Ref.~\cite{yao-tancontact-2018}. This reflects that the generalized Pauli exclusion originates from the competition between repulsive interaction and temperature, manifesting in the length scales $\aOneD$ and $\lambda_T$.

In summary, we report the experimental realization of anyonic thermodynamical ensemble using a 1D quantum gas, manifested through the generalized Pauli exclusion principle. Our approach leverages the intrinsic mapping between the statistical parameter, interparticle interactions and temperature. By systematically measuring the energy-particle relationships, 
we identify equilibrium states quantitatively described by GES, which reveals a continuous crossover between fermionic and bosonic statistics. This framework is further supported by independent probes of pressure and Tan's contact. Our measurements are supported by Yang-Yang thermodynamics, directly connecting microscopic statistical interactions and emergent macroscopic thermodynamic properties. 

Our work opens several promising directions for future research. On such platforms, the emergence of anyonic behavior 
paves a novel path in quantum technologies where effective statistics can be engineered. For instance, it suggests new opportunities for realizing quantum heat engines with anyonic working media~\cite{koch2023quantum,PRXQuantum.2.040312}, which may outperform their bosonic or fermionic counterparts. Also, the unique feature of finite-state occupancy $1/\alpha$ highlights intriguing possibilities for quantum information science~\cite{RevModPhys.80.1083}. By preparing such systems as an array, one may engineer base-N quantum computing and quantum information storage, with $N=1+1/\alpha$. Furthermore, despite their intrinsic differences, GES and FES are closely related~\cite{PRXQuantum.2.040312, Ha1994, PhysRevB.79.064409}. Particles governed by GES exhibit the ability to mimic key thermodynamic properties of topological anyons described by FES~\cite{PRXQuantum.2.040312}. Given the substantial challenges of manipulating and probing the thermodynamics of topological anyons in 2D materials~\cite{feldman2021}, extending the framework developed here may offer practical guidance for their thermodynamic behaviors.

\section*{Acknowledgements}
The authors thank Thierry Giamarchi, Hanns-Christoph N\"agerl, and F. Duncan M. Haldane for their helpful discussions. 
\textbf{Funding:} This work is supported by the National Natural Science Foundation of China (Grant No. 92365208), National Key Research and Development Program of China (Grants No. 2021YFA0718300 and No. 2021YFA1400900), and the Fundamental Research Funds for the Central Universities, Peking University. B.W. is supported by the Fonds de la Recherche Scientifique - FNRS.
\textbf{Author Contributions:}
The work was conceived by H.Y., B.W., F.W. and X.Z. Experiments were prepared and performed by F.W. and D.W. Data were analyzed by F.W., C.Z., and Z.Y. Numerical simulations were performed by C.Z. and Z.Y. Theoretical models were developed by C.Z. and H.Y. The main contributors to the preparation of the manuscript were H.Y., F.W., C.Z., B.W., and X.Z. All authors contributed to the discussion and finalization of the manuscript.
\textbf{Data Availability:} The data shown in the main text is available via Zenodo~\cite{wei_2026_20482260}.
\section*{Method}

\subsection{Experimental setup}
We start with a balanced mixture of the two lowest hyperfine states $|F = 1/2,m_F= \pm 1/2\rangle$ ($|1\rangle$ and $|2\rangle$) of $^{6}\rm Li$ with repulsive interactions in a crossed dipole trap with a wavelength of 1064.5 nm. The waist of each dipole beam is 40 $\mu$m, and the angle between them is $30\degree$. Varying the MOT collection time yields 5,000$\sim$100,000 atoms per spin state. By adjusting the final depth of the optical dipole trap at the end of evaporation cooling, we obtain 3D gases with temperatures ranging from about 10 $\mu$K down to 50 nK, corresponding to a minimum $T/T_{\rm F}\approx 0.1$ with $T_{\rm F}$ the Fermi temperature. Before loading the lattice from the dipole trap, we ramp the magnetic bias field from 780 G to 680 G, which corresponds to an atomic scattering length of 1238.2$a_0$. At this stage, the binding energy between atoms in the two states reaches $\kB \times18.8\rm\ \mu K$, and they pair up to form strongly-coupled Feshbach molecules with a molecular scattering length $a_s=743a_0$. 

Then, the molecules are transformed into 1D tubes of Tonks-Girardeau gases by adiabatically ramping up a 2D triangular optical lattice with a wavelength of 1064 nm within 40 ms. The three lattice beams with vertical polarization are arranged at $120\degree$ relative to each other, each having a waist of 500 $\mu$m. This is much larger than the cloud size, which guarantees the homogeneity of the radial vibrational frequency $\omega_\perp$ across the ensemble. 
We have checked that the tubes are totally uncoupled when $V_{\rm 2D}$ is larger than $4.5 \Er$.
When the lattice depth reaches $10 \Er $, we can achieve an array of uncoupled 1D tubes with a radial frequency $\omega_\perp /2\pi= 6.95$ kHz. The loading process is adiabatic and will change the temperature of the system. Thus, it calls for temperature calibration of 1D systems and we provide a thermometry method below. 
For 1D strongly interacting bosons, the temperature criteria for the Tonks-Girardeau regime reads $T_{\rm TG}=\gamma^2 \hbar^2n^2/2m=4000$ nK~\cite{kheruntsyan2003}. With our temperature estimator, we find the minimum temperature of the 1D system is $T=260$ nK and deeply in this regime with $T/T_{\rm TG}\approx 0.07$. After reaching thermal equilibrium, we turn off the lattice and the dipole trap within several nanoseconds to let the ensemble expand for typically 2.5 ms. The time of flight (TOF) is chosen to be long enough and give the best image quality. Throughout all the stages, the Feshbach bias coils provide a small magnetic field gradient to compensate for the gravitational force. Thus, for measuring the density distribution, we can apply absorption imaging with an image light resonant with $2^2s_{1/2} |1/2,1/2\rangle \to 2^2p_{3/2}|5/2,-1/2\rangle$ in a fixed region of 400 $\mu$m (y) $\times\ 400\ \mu$m (z).


\subsection{Particle number and temperature for 1D system}

The total particle number can be directly extracted from the absorption imaging by an overall integration. Then, we apply a bimodal fitting to the momentum distribution and extract the Thomas-Fermi and Gaussian components. Together with the known trapping frequencies and interactions, we reconstruct the 
3D density profile and project the transverse profile onto the 2D lattice sites, yielding the 1D particle number distribution $N_j$ at given site $j$. According to Refs.~\cite{PhysRevLett.115.085301, Anomalouscooling, Guo2024}, the thermodynamic properties of such tube distributions can be directly mapped to an equivalent single tube with weighted average particle number $N_{\rm 1D} = \sum_j N_{j}^2 / \sum_j N_{j}$. In the Supplementary Information, we run quantum Monte Carlos (QMC) simulations to check whether this single-tube approximation is proper for all the quantities we measured.



Moreover, the temperature of the 1D system is calibrated by two independent methods. On one hand, we perform a standard TOF ballistic expansion and extract the temperature from the width dynamics. On the other hand, following the thermometry methods in Refs.~\cite{gerbier2003, Anomalouscooling, Zhao2025}, we extract the one-body correlation function $G^{(1)}(z)$ by the Fourier transform of the measured momentum distribution $n(k_z)$, and compare it with the QMC simulations under the same conditions at different temperatures. The temperatures obtained from these two methods are consistent with each other. More details are provided in the Supplementary Information.


\subsection{Reduction of free variables for the Lieb-Liniger model}

We show how to reduce the number of free physical parameters of the trapped Lieb-Liniger model~\cite{yao-tancontact-2018}.
Experimentally, any macroscopic thermodynamic observable $\mathcal{A}$ of a trapped 1D Bose gas is determined by four physical parameters: the temperature $T$, chemical potential $\mu$ (equivalently particle number $N_\textrm{1D}$), trap frequency $\omega$ and 1D interaction length $a_{\mathrm{\rm 1D}}$. Thus it can be written as $\mathcal{A}(T, \mu, \omega, \aOneD)$.
We can directly reduce the temperature degree by rescaling the system's energy and length with thermal energy $\kB T$ and 
de Broglie wavelength $\lambda_{\mathrm{T}}$, respectively.
Using dimensional analysis, one can write dimensionless observable $\tilde{\mathcal{A}}$ as a function of three dimensionless parameters $\tilde{\mathcal{A}}(\mu/k_{\mathrm{B}}T, a_{\mathrm{\rm 1D}}/\lambdadB, a_{\mathrm{ho}}/\lambdadB)$. 
Next, assuming the local-density approximation holds, the global observable can be calculated by integrating over local homogeneous slices. Through a spatial coordinate rescaling, $a_{\mathrm{ho}}$ is completely factored out into a global geometric prefactor. More precisely, the core thermodynamic function loses its explicit $\omega$-dependence by 
\begin{equation}
\tilde{A} = \int (\textrm{d} x/\lambdadB)\  \tilde{\mathcal{A}} \left[(\mu-V(x))/\kB T,\aOneD/\lambdadB, a_{\mathrm{ho}}/\lambdadB\right], 
\end{equation}
reducing the description to only two parameters. Finally, fixing the total particle number $N$ eliminates $\mu$ as an independent variable, yielding a universal equation of state governed solely by two dimensionless parameters, defined as the reduced temperature $\xi_T=-\aOneD/\lambdadB$ and the reduced interaction $\xi_{\gamma}=-a_{\rm ho}/\aOneD\sqrt{N_{\rm 1D}}$.

\subsection{Calibration of the effective statistical parameter $\alpha$}

To determine the effective statistical parameter $\alpha$, we compare the thermodynamic properties of the harmonically trapped interacting 1D Bose gas with those of an effective non-interacting gas governed by GES. Specifically, we focus on the energy-particle relation $\varepsilon(N)$ and its derivative $\mathrm{d} \varepsilon / {\mathrm{d} N}$. We calculate this observable as a function of reduced temperature $\xi_T$ and reduced interaction $\xi_\gamma$ using YY-LDA. This yields a two-parameter function $f(\xi_\gamma, \xi_T)$. 
Then, we compute the same observable for a non-interacting GES gas distributed across 1D harmonic oscillator levels~\cite{GES}. This provides a corresponding function $g(\xi_\gamma \xi_T, \alpha)$. Here we assume $\alpha$ to be weakly dependent on quasi-momentum. Because $\alpha$ continuously interpolates between Bose-Einstein ($\alpha=0$) and Fermi-Dirac ($\alpha=1$) statistics, $g(\xi_\gamma \xi_T, \alpha)$ is strictly monotonic with respect to $\alpha$. 
Finally, by equating the exact many-body result with the effective statistical model, $f(\xi_\gamma, \xi_T) = g(\xi_\gamma \xi_T, \alpha)$, we uniquely extract $\alpha$ for any given experimental parameter set ($\xi_\gamma, \xi_T$) via numerical interpolation. Detailed derivations of the scaling forms and numerical procedures are described in the Supplementary Information.

\subsection{Transverse occupancy correction}
In Fig.~\ref{fig:3}b, we find the experimental data approach $\alpha=0$ much faster than the YY-LDA results. We attribute this to the additional occupancy induced by transverse degeneracy.
At low temperature, the condition $\kB T\ll\hbar \omega_\perp$ always holds, and the tubes are frozen at the ground state of the transverse harmonic oscillator.
As $T$ increases, $\kB T$ may reach the scale of $\hbar \omega_\perp$ and excitations along the transverse oscillation level can happen. This induces additional states for the particle to fill in and suppresses the generalized Pauli exclusion. To estimate such an effect, we introduce a transverse degeneracy parameter $D_{\perp}=\kB T/\hbar \omega_\perp$ which is approximately the multiplication factor of the maximum occupation number $f_{max}\sim 1/\alpha$. Then, the effective statistical parameter writes $\tilde{\alpha}=\alpha/D_{\perp}^2$ with the square counting for the two transverse dimensions. We plot $\tilde{\alpha}$ versus $T$ (solid line) in Fig.~\ref{fig:3}b. Although such an approximate calculation is rough, we find that with this correction, the theoretical prediction agrees with the experimental measurement.

\bibliography{biblio-HY}


 \renewcommand{\theequation}{S\arabic{equation}}
 \setcounter{equation}{0}
 \renewcommand{\thefigure}{S\arabic{figure}}
 \setcounter{figure}{0}
 \renewcommand{\thesection}{S\arabic{section}}
 \setcounter{section}{0}
 \onecolumngrid  
     
 
 \newpage
 
{\center \bfseries\large Supplemental Material for \\ \vspace{0.2cm}
\bfseries\large Observation of anyonic thermodynamics and generalized Pauli principle \par}
\vspace{1cm}

In this supplemental material, we provide details about the experimental detections, the effective statistical description of one-dimensional (1D) Lieb-Liniger bosons and the calibration of the effective statistical interaction parameters.

\subsection{The momentum and real space imaging}

\begin{figure*}
    \centering
    \includegraphics[width=0.7\linewidth]{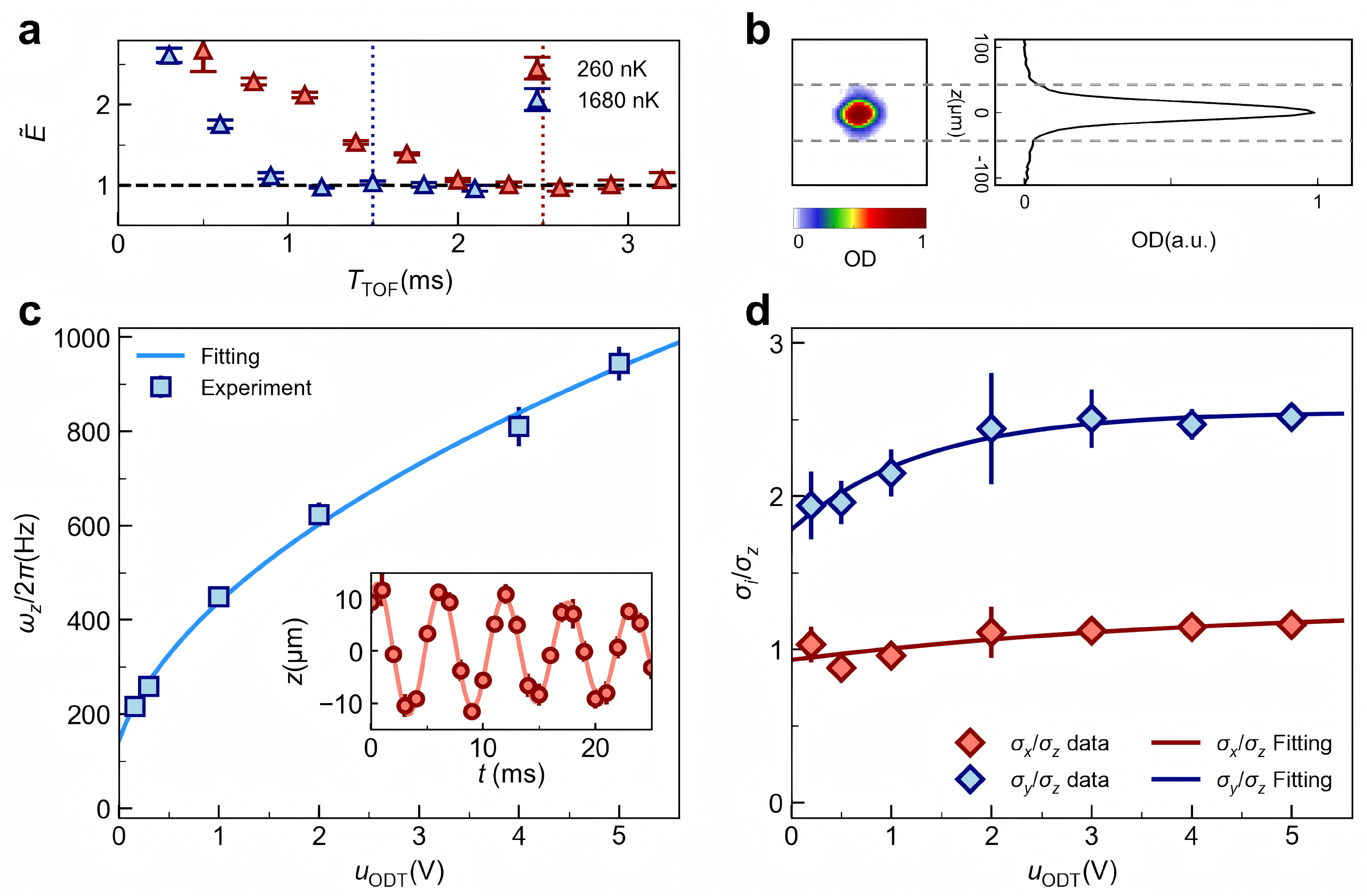}
    \caption{\textbf{Experimental detection and trapping parameters.}
    \textbf{a,} Reduced internal energy versus TOF time for two temperatures $T=260(10)$ nK (scaled by $1.60\times10^4\Er$) and $1680(80)$ nK (scaled by $1.05\times10^5\Er$), at fixed particle number $N_\textrm{total}=5\times10^4$. Dotted lines with corresponding colors show the TOF periods used in the experiment.
    \textbf{b,} Typical \textit{in situ} absorption images for $T = 304(15)$ nK and $a_s = 743a_0$. The right panel shows the obtained density distribution $n(z)$ along the z-direction. 
    \textbf{c,} Trapping frequencies $\omega_z$ for lattice depth $V_{\rm 2D}=10 \Er$ and different ODT control voltages $u_{\rm ODT}$. A typical procedure for measurement via collective-mode excitation is shown in the inset. Red points represent the measured centre-of-mass position along the $z$-axis, and the red line is the corresponding fit. Blue points denote the extracted $\omega_z$ at different $u_{\rm ODT}$, with error bars indicating the fit uncertainties. The blue line is a fit to these data points. 
    \textbf{d,} Aspect ratios $\sigma_i/\sigma _z(i=x,y)$ for lattice depth $V_{\rm 2D}=10 \Er$ and different ODT control voltages $u_{\rm ODT}$. The experimental results of $\sigma_x/\sigma_z$ and $\sigma_y/\sigma_z$ are presented by red and blue dots, respectively, with the correspondingly colored lines representing fits to the data.
    }
    \label{fig:A1}
\end{figure*}

We probe momentum-space information via time-of-flight (TOF) imaging. 
By releasing the molecules from all traps and recording absorption images along the y-axis after free expansion with a long enough duration, we extract $n(k_x,k_z)$. For measurements at different temperatures, we adjusted the TOF duration accordingly to satisfy the far-field condition $t_{\rm TOF}\gg \max (1/\omega_z^2,\sqrt{m\sigma_z^2(0)/\kB T}) $ while maintaining a high signal-to-noise ratio, where $\sigma_z(0)<20\ \rm \mu m$ is the initial half-length of the tubes~\cite{bloch2008}. Then we integrate over $k_x$ and obtain the momentum distribution $n(k_z)$ along the longitudinal direction of the tubes, as shown in Fig.~\ref{fig:2}a2. 
The total internal energy $E$ is obtained by $E=\int n(k_z)k_z^2\mathrm{d}k_z/2m$, together with the total particle number $N_{\rm total}=\int n(k_z)\mathrm{d}k_z$. 
In Fig.~\ref{fig:A1}a, we present the rescaled internal energy $\tilde{E}=E/E_0$ measured at fixed particle number $N_\textrm{total}=5\times10^4$ for two representative temperatures ($T=260(10),1680(80)$ nK) as a function of TOF duration. For these two temperatures, the characteristic energy $E_0$ is taken as $1.60\times10^4\Er$ and $1.05\times10^5\Er$, respectively. Clearly, the extracted internal energy is overestimated in the early stages of expansion. After a sufficiently long TOF, when the cloud size becomes substantially larger than its initial in-trap size, the measured energy saturates at a stable value (black dashed line). For even longer expansion times, the signal-to-noise ratio decreases due to reduced density of the cloud, leading to increased measurement uncertainty. Accordingly, we characterized the expansion dynamics for each temperature and selected a TOF (dotted lines) that balances these two effects, ensuring both the far-field condition and a reliable signal-to-noise ratio. 

Also, we measure the real-space information via \textit{in situ} absorption imaging of the trapped particles. Thanks to the light mass of the lithium molecules, the particle density is low enough that it enables the macroscopic density distribution to be resolved without a high spatial resolution requirement. Following a similar procedure as for the momentum distribution, we integrate the density distribution along the $x$-axis to obtain the longitudinal density profile $n(z)$ along the tube, from which we extract the pressure. In Fig.~\ref{fig:A1}b, we show one example for the case of $T = 304(15)$ nK, $V_{\rm 2D}=10\Er$ and $a_s = 743a_0$. The gray dashed lines mark the maximum cutoff of the density distribution, beyond which noise becomes the dominant contribution. 
The pressure of the system can be directly obtained by integrating this data.
\subsection{Trap frequency and aspect ratio}

We calibrate the trap frequency by \textit{in situ} imaging of oscillation dynamics. In the experiment, the temperature is tuned via the control voltages $u_{\rm ODT}$ of the optical dipole trap (ODT), and the system's trap frequency arises from the joint contribution of the ODT and the optical lattice. To determine the axial trap frequency $\omega _z$ under different $u_{\rm ODT}$, we excite the dipole mode by a perturbation of the ODT and monitor the resulting oscillation of the centre-of-mass position of the density distribution~\cite{PhysRevLett.99.150403}. The measured frequencies and a demonstration of oscillation with $V_{\rm 2D}=10\Er$ and $u_{\rm ODT}=0$ are shown in Fig.~\ref{fig:A1}c and its inset, respectively. The dynamics of the centre-of-mass position are shown as red circles and fitted by the red oscillatory curve to extract the frequency information. The extracted frequencies at different $u_{\rm ODT}$ (blue squares) are then fitted with the function $f(u)=b_1\cdot\sqrt{u+b_2}$ (blue line), where $b_1,b_2$ are fitting parameters, providing the trap frequency for all voltages $u_{\rm ODT}$. This calibration is used in all subsequent theoretical calculations.

Then, the aspect ratio is determined by probing the two-dimensional density distribution along two orthogonal directions. We uniformly adopt a Gaussian fit as the standard for extracting the characteristic lengths along different dimensions $\sigma_i\ (i=x,y,z)$. The \textit{in situ} imaging along $y$- and $z$- axis yields aspect ratio $\sigma_x/\sigma _z$ and $\sigma_y/\sigma _x$, correspondingly. These results are presented in Fig.~\ref{fig:A1}d by red ($\sigma_x/\sigma _z$) and blue ($\sigma_y/\sigma _z$) dots. Within the range of $u_{\rm ODT}$ explored in this experiment, $\sigma_x/\sigma _z$ remains close to unity, whereas $\sigma_y/\sigma _z$ increases monotonically and significantly with increasing $u_{\rm ODT}$. 
As $u_{\rm ODT}$ increases, the overall trapping potential becomes increasingly dominated by the optical dipole trap. Consequently, the aspect ratio of the gas approaches that of the ODT itself. The ODT is elongated along the $y$-axis (radial), with an intrinsic radial-to-axial ratio of approximately 2.5. This behaviour is fully consistent with our experimental observations, confirming the experimental geometry.
We fit the aspect ratio by $r(u)=c_1-c_2\cdot e^{-c_3u}$ where $c_1,c_2,c_3$ are fitting parameters (solid lines), and use these fits as the scaling basis for reconstructing the three-dimensional density distribution. This is subsequently used to determine the number of tubes and the equivalent 1D particle number $N_{\rm 1D}$.

\subsection{Determination of the Density Profiles}\label{S1:C}

Our experimental setup contains 2D arrays of 1D tubes. In the manuscript, we treat our measurement as an equivalent 1D tubes with a weighted number of particles 
\begin{align}
\label{eq:Weighted}
N_{\rm 1D} = \sum_j N_{j}^2 / \sum_j N_{j},
\end{align}
where $N_{j}$ denotes the particle number in the $j$-th tube. 
According to Refs.~\cite{PhysRevLett.115.085301, Anomalouscooling, Guo2024, science.abn1719},
this mapping is correct for measuring the momentum distribution $n(k)$ while the particle number distribution follows a Thomas-Fermi profile. Here, we further verify the determination of the particle number distribution in our experiment and the validity of such an equivalent 1D tube description.
 
Since we prepare the system under different temperatures, we cannot guarantee a Thomas-Fermi distribution for the 3D density profile before loading the 2D lattices. Therefore, we perform a bimodal fit to the measured momentum distribution and obtain the portions of Thomas-Fermi (TF, $N_{\text{TF}}$) and Gaussian ($N_{\text{G}}$) components. 
For each component, we reconstruct its transverse bimodal density profile in the $xy$ plane. Then, we obtain the particle number distribution by mapping this profile onto the discrete 2D lattice sites. 
For the TF component, the two-dimensional profile is characterized by the transverse radii 
\begin{align}
R_i= \sqrt{\frac{2\mu'}{m\omega_i^2}}, \ (i=x,y,z),
\end{align}
where the modified chemical potential is given by 
\begin{align}
\mu' = \frac{\hbar \bar{\omega}}{2} \left( 15 N_\textrm{total} \frac{a_s}{\bar{a}_\textrm{ho}} \frac{\tilde{g}}{g} \right)^{2/5},
\end{align}
with $\bar{\omega}=\sqrt[3]{\omega_x\omega_y\omega_z}$ and $\bar{a}_\textrm{ho}=\sqrt{\hbar/m\bar{\omega}}$. Here, considering the time scale of the loading process, we rescale the coupling constant by~\cite{tenart:tel-03560800} 
\begin{align}
\tilde{g} = g \frac{\pi(V_\textrm{2D}/\Er)^{1/2}}{2 \left( \text{Erf} \left[ \pi(V_\textrm{2D}/\Er)^{1/4}/2 \right] \right)^2}. 
\end{align}
The TF particle number in the $j$-th tube $N_{j,\textrm{TF}}$ is then obtained with $\mu'$ and $\tilde{g}$. For the Gaussian component, the fitted $\sigma_z$ together with the aspect ratios from Fig.~\ref{fig:A1}d yields the transverse Gaussian widths $\sigma_x$ and $\sigma_y$, allowing us to determine the Gaussian particle number $N_{j,\textrm{G}}$ in the $j$-th tube. Summing up the two contributions gives the particle number in the $j$-th tube $N_j=N_{j,\textrm{TF}}+N_{j,\textrm{G}}$. Finally, we obtain the weighted average particle number $N_{\rm 1D}$ by Eq.~(\ref{eq:Weighted}). 

\begin{figure*}
    \centering
    \includegraphics[width=0.95\linewidth]{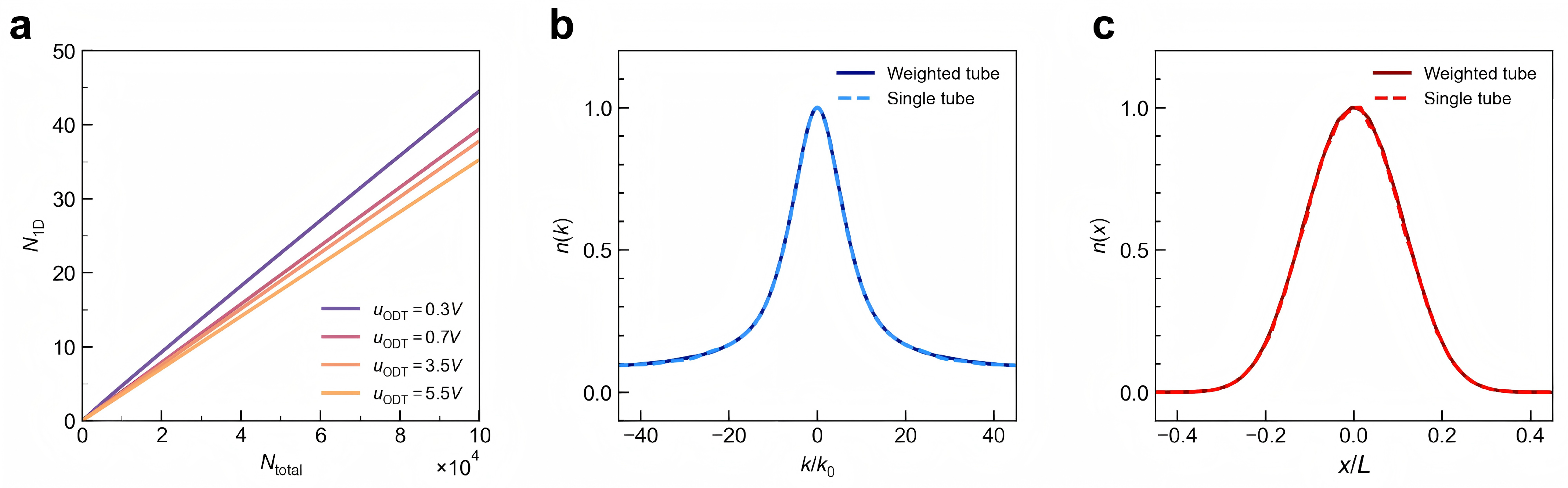}
    \caption{\textbf{Calculation of the effective 1D particle number.} 
    \textbf{a,} The calculated $N_{\rm 1D}-N_{\rm total}$ mapping for $u_{\rm ODT}=0.3, 0.7,3.5,5.5$ V, with $V_\textrm{2D}=10\Er$ and $a_s = 743a_0$.
    \textbf{b-c,} The QMC results of 1D momentum distribution (\textbf{b}) and density distribution (\textbf{c}) with $N_{\rm total}=4 \times 10^4$, $T = 1400\text{ nK}$ and $\omega_z/2\pi =378\text{ Hz}$. The dark solid line and the light dashed line represent the results for the weighted tube array and a single tube, respectively.
    }
    \label{fig:A2}
\end{figure*}

In this way, we obtain the particle number for a weighted average tube $N_{\rm 1D}$ as a function of the total particle number $N_{\rm total}$ under different values of $u_{\rm ODT}$.
In Fig.~\ref{fig:A2}a, we show $N_{\rm 1D}$ versus $N_{\rm total}$ at several representative ODT voltages. The slope $\textrm{d}N_{\rm 1D}/\textrm{d}N_{\rm total}$ decreases systematically with increasing voltage. This is consistent with the increase of $\sigma_x$ and $\sigma_y$ (equivalently an increasing number of tubes) while increasing $u_{\rm ODT}$ in Fig.~\ref{fig:A1}d.

To further verify the validity of the weighted-average treatment, we perform numerical simulations based on the path-integral Monte Carlo (PIMC) method~\cite{ceperley1995path}. 
We simulated the system with the Hamiltonian in Eq. (3)
of the main text. Under the grand-canonical ensemble, we compute the density distribution $n(z)$ by counting the real-space worldlines.
Thanks to the worm-algorithm implementations~\cite{boninsegni-worm-short-2006, boninsegni-worm-long-2006}, the simulations are extended efficiently to the open worldline configuration space, where the statistics of creation and annihilation operators with open ends are counted to compute the $z$-axis momentum distribution $n(k_z)$ in the open worldline configuration.
We run the QMC simulation for the tube array at $N_{\rm total}=4 \times 10^4$,  $T = 260\text{ nK}$ and $\omega_z/2\pi =213\text{ Hz}$ with the same particle number distribution as in our experiment, and compare it with those of a single tube containing the equivalent 1D particle number $N_\textrm{1D}=15$ under the same conditions.
The normalized density and momentum distributions are presented in Fig.~\ref{fig:A2}b and Fig.~\ref{fig:A2}c, respectively. The single-tube results (dashed lines) are found to be almost perfectly coincident with those of the weighted tube array (solid lines). Therefore, we confirm that the weighted-average particle number method is fair to describe our experimental system in both real and momentum spaces.

\begin{figure*}
    \centering
    \includegraphics[width=0.9\linewidth]{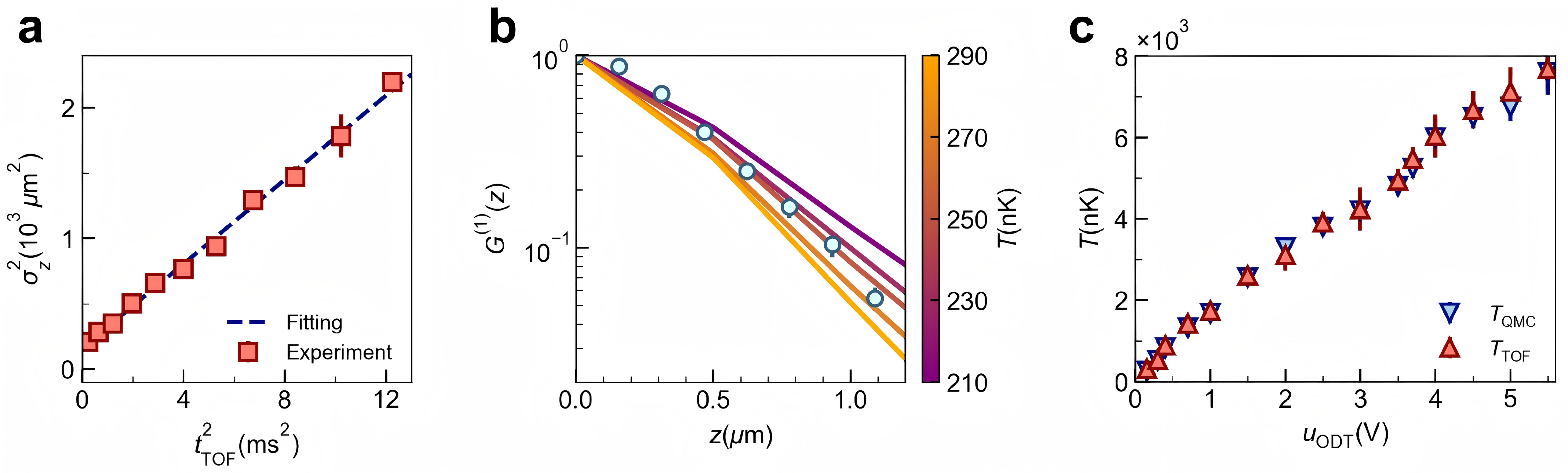}
    \caption{\textbf{Temperature calibration. }
    \textbf{a,} Calibration by the TOF method for $u_{\rm ODT}=0.16$ and $V_\textrm{2D}=10\Er$. The red squares and blue line are experimental data and their corresponding fitting line, respectively. 
    \textbf{b,} Calibration by the QMC method for $u_{\rm ODT}=0.16$ and $V_\textrm{2D}=10\Er$. Blue circles denote $G^{(1)}$ extracted from the experimental data, while the lines represent the QMC results at five equally spaced temperatures indicated by the color bar.
    \textbf{c,} Temperature results of two different methods. The results of the TOF method and the QMC method are marked by red and blue dots, respectively.
    }
    \label{fig:A3}
\end{figure*}

\subsection{The Temperature Calibration}

After loading the 2D optical lattices adiabatically, the temperature of the system will change. 
We determine the temperature of our experimental 1D system via two independent methods: TOF ballistic expansion calibration (TOF method) and quantum Monte Carlo thermometry (QMC method).

The TOF method relies on the free expansion of the thermal component after being released from the trap, which is a commonly used method for temperature calibration~\cite{science2695221198, bloch2008}. The squared Gaussian width of the cloud $\sigma^2(t)$ evolves with the expansion time $t$ as
\begin{align}
\sigma^2(t) = \sigma_0^2 + \frac{k_B T}{m} t_{\rm TOF}^2,
\end{align}
where $\sigma_0$ is the initial Gaussian width in the trap. A typical measurement for $u_{\rm ODT}=0.16$, $V_\textrm{2D}=10\Er$ and $a_s = 743a_0$, with $t_\textrm{TOF}$ ranging from 0.5 ms to 3.5 ms, is shown in Fig.~\ref{fig:A3}a. By linearly fitting the measured $\sigma_z^2$ as a function of $t_\textrm{TOF}^2$, the temperature $T_\textrm{TOF}=233(10)$ nK is extracted directly from the slope.

For low dimensional quantum gases, the QMC method is a commonly used approach for thermometry~\cite{gerbier2003,Anomalouscooling,Zhao2025}, since the decay of one-body correlation function $g^{(1)}\left(z, z^{\prime}\right)=\langle\hat{\Psi}^{\dagger}\left(z^{\prime}\right) \hat{\Psi}(z)\rangle$ has a significant temperature dependence. 
Experimentally, the integrated correlation function can be obtained via a Fourier transform of the measured momentum distribution $n(k_z)$, yielded by
\begin{align}
G^{(1)}(z)=\iint \textrm{d} z^{\prime} g^{(1)}\left(z^{\prime}+z, z^{\prime}\right).
\end{align}
In practice, we perform QMC simulations of 
$G^{(1)}_{\mathrm{QMC}}(z)$ at various temperatures $T$ and comparing with the experimental data.
In Fig.~\ref{fig:A3}b, we show the example for the case $u_{\rm ODT}=0.16$, $V_\textrm{2D}=10\Er$ and $a_s = 743a_0$.
The experimental data are presented by blue circles, and the QMC simulations at $T=210,230,250,270,290$ nK are plotted as colored solid lines which serve as a ruler.
Comparing the two allows us to determine the experimental temperature $T_\textrm{QMC}=260(10)$ nK.


The temperatures at different control voltages $u_{\rm ODT}$ extracted from these two different methods are shown in Fig.~\ref{fig:A3}c. The two sets of results are in excellent agreement within the error bars. Based on previous experiments~\cite{science2695221198, bloch2008, wurz2017, Anomalouscooling, Zhao2025}, the QMC and TOF methods are more reliable at low and high temperatures, respectively. Our calibration results confirm that each method yields smaller uncertainties within its respective favorable regimes. Therefore, we use the QMC-based temperatures for $T<2000$ nK and the TOF-based temperatures for $T>2000$ nK.

\section{Effective Statistical Description of Uniform 1D Interacting Bosons}
\label{sec:appendix_GES}

In this section, we establish the thermodynamic equivalence between the uniform 1D Lieb-Liniger gas and an ideal gas obeying Haldane's Generalized Exclusion Statistics (GES)~\cite{Wu_notes, Guan1, Guan2, Guan3}. By analyzing the many-body eigenstates in quasi-momentum space, we construct a microscopic statistical picture based on the repulsion behavior of the Bethe roots.

Our statistical mapping is fundamentally grounded in the exact microscopic solutions of the 1D Lieb-Liniger model~\cite{lieb1963a}. For the repulsive Lieb-Liniger gas, the many-body eigenstates are characterized by a set of quasi-momenta $\{k_j\}$. Subject to periodic boundary conditions, these quasi-momenta satisfy the transcendental Bethe Ansatz equations~\cite{lieb1963a}
\begin{equation}
    \label{eq:discreteTBA}
    k_j L = 2\pi I_j + 2\sum_{l \neq j}^{N_{\rm{1D}}} \arctan\left(\frac{k_j-k_l}{\tilde{g}_{\rm 1D}}\right), \quad j=1,\cdots,N_{1D},
\end{equation}
where $L$ is the system length and $\tilde{g}_{\rm 1D} \equiv m g_{\rm 1D}/\hbar^2$ is the rescaled interaction parameter with the dimension of momentum. This parameter connects to the Lieb-Liniger parameter $\gamma$ via $\tilde{g}_{\rm 1D} = n\gamma$, with the 1D density $n = N_{\rm{1D}}/L$. Here, $N_{\rm{1D}}$ denotes the particle number in a single tube, as in the main text. $\{I_j\}$ is a set of distinct integers (odd $N_{\rm 1D}$) or half-integers (even $N_{\rm 1D}$). The second term on the right-hand side of Eq.~(\ref{eq:discreteTBA}) encapsulates the two-body scattering phase shift. Notably, the requirement of mutually distinct quantum numbers $I_j$ implies that, at the level of state counting, the bosonic Bethe roots exhibit Pauli-like exclusion.

In the thermodynamic limit ($N_{\rm 1D}, L \to \infty$ with $n$ finite), the Bethe roots form a dense continuous distribution. Figure.~\ref{fig.S4}(a) visualizes its zero-temperature profile. As depicted, in the Tonks-Girardeau limit ($\gamma \rightarrow \infty$), the scaled root density $2\pi\rho(k)$ presents a standard free Fermi sea with a normalized height of unity and a sharp cutoff at the Fermi momentum $k_F=\pi n$. In the strong interaction regime ($\gamma \gg 1$), this distribution schematically deforms into a denser plateau characterized by an enhanced density height of $1/\alpha$ and a narrowed cutoff momentum of $\alpha k_F$. Here, $\alpha$ is the statistical parameter defined by Eqs.~(\ref{Eq:GES1}) and (\ref{Eq:GES2}) 
in the main text, which will be analytically derived later in this section. In the weak interaction regime ($\gamma\ll 1$), the state exclusion effect diminishes and the roots condense into a narrow cluster near $k=0$.

To connect these microscopic Bethe roots to macroscopic thermodynamics, we utilize the standard Yang-Yang formalism~\cite{yang1969}. A continuous counting function $I(k)$ can be defined as an interpolation of the quantum numbers $\{I_j\}$ in Eq.~(\ref{eq:discreteTBA}) given by
\begin{equation}
    I(k) = \frac{L}{2\pi}k + \frac{1}{\pi}\sum_{l=1}^{N_{\rm{1D}}}\arctan\left(\frac{k-k_l}{\tilde{g}_{\rm 1D}}\right).
\end{equation}
Defining the total density of states as $\rho_t(k) = \lim_{\Delta k\rightarrow 0} \Delta I/L\Delta k$, one obtains the Yang-Yang equation. It yields the total available density of states at momentum $k$ 
\begin{equation}
    \label{eq:YY}
    \rho_t(k)\equiv \rho(k) + \rho_h(k)  = \frac{1}{2\pi} + \frac{1}{2\pi}\int_{-\infty}^{\infty}\frac{2\tilde{g}_{\rm 1D}}{\tilde{g}_{\rm 1D}^2+(k-k')^2}\rho(k')\textrm{d}k',
\end{equation}
where $\rho(k)$ is the occupied particle density and $\rho_h(k)$ is the unoccupied holes density. This equation demonstrates that the scattering phase shift dynamically reshapes the state space. The number of available states at $k$ is depleted by particle occupations at other momenta $k'$. At zero temperature, the holes vanish ($\rho_h=0$) within the Fermi sea, and the integration range simply truncates to $[-k_F, k_F]$.

At finite temperatures, the thermal equilibrium state is determined by minimizing the grand potential functional $\Omega = E - \mu N_{\rm 1D} - TS$ subject to the kinematic constraint of Eq.~(\ref{eq:YY}), with $S$ the entropy. By expressing the entropy in terms of $\rho(k)$ and $\rho_h(k)$ and applying the variational condition $\delta\Omega = 0$, one introduces the dressed energy $\epsilon(k)$ defined via $\rho_h(k)/\rho(k) = e^{\epsilon(k)/\kB T}$. Through this variational approach, one directly obtains the exact macroscopic equation of state and the self-consistent dressed energy equation~\cite{yang1969}
\begin{align}
    \label{eq:YY_Omega}
    \frac{\Omega}{L} &= -\kB T \int_{-\infty}^{\infty} \frac{\textrm{d}q}{2\pi} \ln{\left[1+e^{-\frac{\epsilon(q)}{\kB T}}\right]}, \\
    \label{eq:YY_epsilon}
    \epsilon(k) &= \frac{\hbar^2 k^2}{2m} - \mu - \kB T \int_{-\infty}^{\infty} \frac{\textrm{d}q}{2\pi} \frac{2\tilde{g}_{\rm 1D}}{\tilde{g}_{\rm 1D}^2+(k-q)^2} \ln{\left[1+e^{-\frac{\epsilon(q)}{\kB T}}\right]}.
\end{align}
To avoid ambiguity, we emphasize that the microscopic dressed energy $\epsilon(k)$ is distinct from the macroscopic ensemble-averaged per-particle energy $\varepsilon$ defined in the main text.

Having established the continuous picture, we now recall the formal mapping onto Haldane's GES formalism~\cite{Guan1, Guan2, Guan3}. Specifically, a variation $\Delta N_j$ in the particle occupation of state $j$ induces a change $\Delta d_i$ in the available holes in state $i$, given by $\Delta d_i = -\sum_{j}\alpha_{ij}\Delta N_j$. In continuous quasi-momentum space, the effective hole density is simply $\rho_h(k)$, and the exclusion relation becomes
\begin{equation}
    \Delta \rho_h(k) = -\int_{-\infty}^{\infty} \alpha(k,k') \Delta \rho(k') \textrm{d}k'.
\end{equation}
By comparing this with the differential form of Eq.~(\ref{eq:YY}), the equivalent exclusion parameter matrix is identified as
\begin{equation}
    \label{eq:mutual_alpha}
    \alpha(k,k') = \delta(k-k') - \frac{1}{\pi}\frac{\tilde{g}_{\rm 1D}}{\tilde{g}_{\rm 1D}^2+(k-k')^2}.
\end{equation}
The delta function reflects the intrinsic fermionic nature of the Bethe roots in state counting, while the Lorentzian term provides the statistical correction originating from dynamical collisions. Thus, the dynamical interaction in real space is exactly mapped to a mutual statistical interaction in quasi-momentum space.

To obtain a practical scalar parameter for our phenomenological model, we can simplify the exact mutual matrix Eq.~(\ref{eq:mutual_alpha}) under the strong interaction ($\gamma \gg 1$) and low-temperature limit. It naturally reduces to  $\alpha(k,k') \approx \alpha \delta(k-k')$. Integrating $\alpha(k,k')$ over the background Fermi sea $[-\pi n, \pi n]$ yields the effective momentum-dependent parameter
\begin{equation}
    \alpha(k) = \int_{-\pi n}^{\pi n} \alpha(k,k') \textrm{d}k' = 1 - \frac{1}{\pi}\left[\arctan\left(\frac{k+\pi n}{\tilde{g}_{\rm 1D}}\right) - \arctan\left(\frac{k-\pi n}{\tilde{g}_{\rm 1D}}\right)\right].
\end{equation}
For low-energy excitations near the Fermi surface ($k \sim k_F \ll \tilde{g}_{\rm 1D}$), $\alpha(k)$ asymptotically approaches a constant. Substituting $\tilde{g}_{\rm 1D} = n\gamma$, one can extract the non-mutual statistical parameter as $\alpha = 1 - \frac{2}{\gamma} + \mathcal{O}(\gamma^{-2})$.
This recovers the result of Ref.~\cite{Guan1, Guan2, Guan3}.  Consequently, the modified plateau height ($1/\alpha$) and cutoff momentum ($\alpha k_F$) in the root density distribution (Fig.~\ref{fig.S4}(a)) are direct manifestations of this extracted statistical parameter in the strong interacting regime. At zero temperature under strong repulsion, the macroscopic state-counting is accurately described by a single-parameter non-mutual GES. At finite temperatures, thermal excitations can conquer the repulsive interaction and allow real-space overlap, which suppresses the fermionization process. Thus, an equivalent GES statistic can appear.
In this manuscript, we assume a non-mutual $\alpha$ through our analysis. Since we extract $\alpha$ from the macroscopic observable $\partial\varepsilon/\partial N_{\rm 1D}$, it naturally serves as a thermally averaged effective parameter. This approach provides a simplified phenomenological description of the bulk equation of state.


\section{Calibration of the Effective Statistical Interaction Parameter}
\label{calibration}

In this section, we outline the numerical calibration scheme to establish a quantitative bridge between microscopic interactions and macroscopic effective statistics. Our primary goal is to uniquely determine the effective exclusion parameter $\alpha$ for a harmonically trapped 1D Bose gas at finite temperatures. To achieve this, we construct a thermodynamic mapping by matching the exact Yang-Yang thermodynamics of the fully interacting system to the simplified non-interacting GES model.

As established in the Methods, applying dimensional analysis and the local-density approximation (LDA) reduces the thermodynamics of the trapped Lieb-Liniger gas to a two-parameter scaling form~\cite{yao-tancontact-2018}. Specifically, the grand potential $\Omega$ takes the form
\begin{align}
    \frac{\Omega}{\kB  T} = \left(\frac{a_{\rm ho}}{\lambda_T}\right)^2 \mathcal{A}_\Omega\left(\frac{\mu}{\kB  T}, \frac{\aOneD}{\lambda_T}\right),
\end{align}
with $\mathcal{A}_\Omega$ being a dimensionless scaling function. Using the standard thermodynamic relations for the particle number $N_{\rm 1D}=-\partial\Omega/\partial\mu|_{T,\aOneD}$, the entropy $S=-\partial\Omega/\partial T|_{\mu,\aOneD}$, and the energy $E=\Omega+TS+\mu N_{\rm 1D}$, the ensemble-averaged energy per particle $\varepsilon$ can be expressed as a universal function of the reduced temperature $\xi_T$ and the reduced interaction $\xi_\gamma$, \ie\ $\varepsilon/\kB T = \mathcal{A}_\varepsilon^*(\xi_\gamma, \xi_T).$
Consequently, the variation of the per-particle energy with respect to particle number $N_{\rm 1D}$ is given by
\begin{equation}
    \label{eq:YY_scaling}
    \frac{1}{\hbar\omega} \frac{\partial \varepsilon}{\partial N_{\rm 1D}} = f(\xi_\gamma, \xi_T),
\end{equation}
with $f$ being the dimensionless target function derived from the exact YY-LDA solution.

\begin{figure*}
    \centering
    \includegraphics[width=0.8\linewidth]{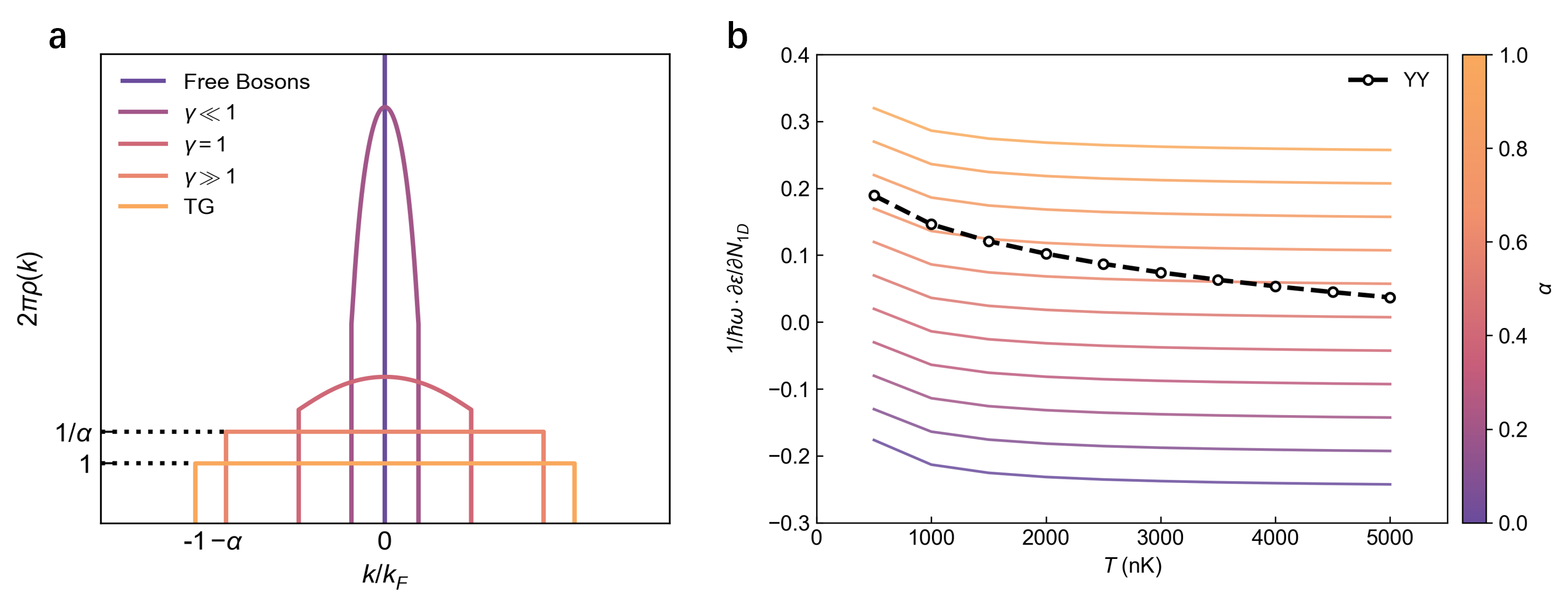} 
    \caption{\textbf{Manifestation and numerical calibration of the effective statistical parameter $\alpha$.}
    \textbf{a}, The scaled Bethe root density $2\pi\rho(k)$ versus the dimensionless momentum $k/k_F$ at $T=0$. The five curves represent different interaction regimes: the Tonks-Girardeau (TG) limit, strong interaction ($\gamma \gg 1$), intermediate interaction ($\gamma = 1$), weak interaction ($\gamma \ll 1$), and free bosons. In the TG limit ($\gamma \rightarrow \infty$), the distribution is identical to a free Fermi sea with a normalized height of 1 and a cutoff momentum of $k_F$. In the strong interaction regime ($\gamma \gg 1$), governed by the generalized exclusion statistics (GES), the roots form a denser plateau with a narrowed cutoff momentum $\alpha k_F$ and an enhanced density height $1/\alpha$. 
    \textbf{b}, Numerical calibration of $\alpha$ from finite-temperature thermodynamics. The variation $\partial\varepsilon/\partial N_{\rm{1D}}$ extracted via linear fits for $N_{\rm 1D}$ ranging from 5 to 40 is plotted against temperature $T$. The solid colored lines represent the non-interacting GES results for constant $\alpha$ values ranging from 0 to 1 in steps of 0.1. The exact YY-LDA calculation for an interacting 1D Bose gas of $^6$Li molecules is overlaid as a black dashed line with open circles.
    By matching this exact thermodynamic trajectory onto the monotonic background GES contours, the effective parameter $\alpha$ is uniquely extracted via numerical interpolation. For this illustrative example, the parameters are set to a trap frequency $\omega = 2\pi\times 500$~Hz and a coupling constant $g_{\rm{1D}}=7.4317\times 10^{-36} $~J$\cdot$m.}
    \label{fig.S4}
\end{figure*}

Conversely, within the GES framework, the uniform Lieb-Liniger gas is mapped to a non-interacting gas obeying fractional exclusion statistics~\cite{GES,Guan1,Guan2,Guan3}. Starting from the exact GES statistical distribution given by Eqs.~(\ref{Eq:GES1}) and (\ref{Eq:GES2}) 
of the main text, and following a dimensional analysis analogous to Ref.~\cite{yao-tancontact-2018}, the dimensionless grand potential of a uniform gas takes the scaling form $\Omega/(k_{\rm B}T) = (L/\lambda_T) \mathcal{G}_h(\mu/\kB T, \alpha)$. For the harmonically trapped gas, we apply the local-density approximation (LDA) by integrating the local homogeneous contributions over the trap potential $V(z) = \frac{1}{2}m\omega^2 z^2$. Strictly speaking, the effective parameter $\alpha$ is a locally varying quantity dependent on the spatial density profile. However, to preserve the analytical scaling properties and extract a macroscopic equation of state, we adopt a global, thermally averaged statistical parameter $\alpha$ as a practical approximation. Under this assumption, the total grand potential is given by
\begin{equation}
    \label{eq:GES_LDA_integral}
    \frac{\Omega}{k_{\rm B}T} = \int \frac{\textrm{d}z}{\lambda_T} \, \mathcal{G}_{h}\left(\frac{\mu - V(z)}{k_{\rm B}T}, \alpha\right).
\end{equation}
By rescaling the spatial coordinate $z$ with the characteristic length scale $a_{\rm ho}^2/\lambda_T$, the trap potential naturally introduces the macroscopic energy scale $\hbar\omega$. The grand potential then reduces to a two-parameter scaling form
\begin{equation}
    \label{eq:GES_trap_scaling}
    \frac{\Omega}{k_{\rm B}T} = \left(\frac{a_{\rm ho}}{\lambda_T}\right)^2 \mathcal{G}^*\left(\frac{\mu}{k_{\rm B}T}, \alpha\right),
\end{equation}
with $\mathcal{G}$ being a dimensionless integration function. Consequently, the thermodynamic relation for the total particle number, $N_{\rm 1D} = -\partial\Omega/\partial\mu|_{T,\alpha}$, dictates that the reduced chemical potential $\mu/k_{\rm B}T$ is universally determined by $\alpha$ and the specific combination $N_{\rm 1D}(\lambda_T/a_{\rm ho})^2$. Recognizing that $N_{\rm 1D}(\lambda_T/a_{\rm ho})^2 \propto (\xi_\gamma \xi_T)^{-2}$, a parallel derivation to the YY-LDA solution yields the dimensionless derivative of the per-particle energy with respect to $N_{\rm 1D}$ as
\begin{equation}
    \label{eq:GES_scaling}
    \frac{1}{\hbar\omega} \frac{\partial \varepsilon}{\partial N_{\rm 1D}} = g(\xi_\gamma \xi_T, \alpha).
\end{equation}
Crucially, the 1D scattering length $\aOneD$ explicitly cancels out in the product $\xi_\gamma \xi_T = a_{\rm ho}/(\lambda_T\sqrt{N_{\rm 1D}})$, structurally verifying that the GES formulation effectively describes a non-interacting system parameterized solely by $\alpha$.

Equating Eq.~(\ref{eq:YY_scaling}) and Eq.~(\ref{eq:GES_scaling}) enforces the macroscopic thermodynamic equivalence between the microscopic model and the phenomenological GES description
\begin{equation}
    \label{eq:equivalence}
    f(\xi_\gamma, \xi_T) = g(\xi_\gamma \xi_T, \alpha).
\end{equation}
Because the function $g$ monotonically maps the degree of Pauli exclusion (from $\alpha=0$ for bosons to $\alpha=1$ for fermions), the effective statistical parameter $\alpha$ is uniquely determined via numerical inversion as
\begin{equation}
    \label{eq:final_mapping}
    \alpha = h(\xi_\gamma, \xi_T) = g^{-1}\big(\xi_\gamma \xi_T, f(\xi_\gamma, \xi_T)\big).
\end{equation}

The exact thermodynamic target function $f(\xi_\gamma, \xi_T)$ for the trapped Lieb-Liniger gas is computed by applying LDA to the homogeneous Yang-Yang thermodynamic equations in Eqs.~(\ref{eq:YY_Omega}) and (\ref{eq:YY_epsilon}). For the non-interacting GES system, we directly evaluate $g(\xi_\gamma \xi_T, \alpha)$ by applying the exact statistical distribution [Eqs.~(\ref{Eq:GES1}) and (\ref{Eq:GES2}) in the main text]
to the 1D harmonic oscillator levels, $E_j = (j+1/2)\hbar\omega$.

In practice, for a given set of system parameters $(\xi_\gamma, \xi_T)$, we first compute the exact thermodynamic target $f(\xi_\gamma, \xi_T)$ using the Yang-Yang formalism. We then generate a set of reference curves by evaluating the non-interacting GES function $g(\xi_\gamma \xi_T, \alpha)$ across a discrete domain of $\alpha \in [0,1]$ in steps of 0.1.
To demonstrate this procedure, we present a specific example of an interacting 1D Bose gas of $^6$Li molecules. Here, the coupling constant is set to $g_{\rm{1D}} = 7.4317\times 10^{-36}$~J$\cdot$m, to be consistent with our experiment, while the trap frequency is chosen as $\omega = 2\pi\times 500$~Hz to represent a typical value within the experimental parameter range. As illustrated in Fig.~\ref{fig.S4}(b), plotting the exact Yang-Yang results for this system reveals that the interacting trajectory smoothly crosses the background GES reference contours. By matching the exact values onto these contours, the effective exclusion parameter $\alpha$ is uniquely extracted via numerical interpolation. From this mapping procedure, we conclude that the finite-temperature fermionization of the interacting Bose gas can be systematically and quantitatively captured by a single emergent statistical parameter $\alpha$.

\end{document}